\documentclass[12pt, letterpaper]{article}
\usepackage[margin=1in]{geometry} 

\usepackage{graphicx}
\usepackage{color}
\usepackage{float}
\usepackage{amsmath}
\usepackage{amssymb}
\usepackage{bm}
\usepackage{setspace}
\usepackage{caption}
\usepackage{subcaption}
\usepackage{titlesec}
\usepackage{xr-hyper}
\usepackage{hyperref}
\usepackage{authblk}
\usepackage{multirow}

\usepackage{filecontents}

\DeclareMathOperator{\E}{\mathbb{E}}

\makeatletter
\newcommand*{\addFileDependency}[1]{
  \typeout{(#1)}
  \@addtofilelist{#1}
  \IfFileExists{#1}{}{\typeout{No file #1.}}
}
\makeatother
 
\newcommand*{\myexternaldocument}[1]{%
    \externaldocument{#1}%
    \addFileDependency{#1.tex}%
    \addFileDependency{#1.aux}%
}

\myexternaldocument{SI}

\providecommand{\keywords}[1]{\textbf{keywords:} #1}

\title{Three Dimensional Cluster Analysis for Atom Probe Tomography Using Ripley's K-function and Machine Learning}

\author[a, b]{Galen B. Vincent}
\author[b]{Andrew P. Proudian}
\author[b]{Jeramy D. Zimmerman \thanks{Corresponding Author. Email: jdzimmer@mines.edu}}

\affil[a]{Department of Applied Mathematics and Statistics, Colorado School of Mines, Golden, CO 80401, USA}
\affil[b]{Department of Physics, Colorado School of Mines, Golden, CO 80401, USA}

\begin{document}
\maketitle
\thispagestyle{empty}
\pagestyle{plain}

\begin{abstract}
The size and structure of spatial molecular and atomic clustering can significantly impact material properties and is therefore important to accurately quantify. Ripley's K-function ($K(r)$), a measure of spatial correlation, can be used to perform such quantification when the material system of interest can be represented as a marked point pattern. This work demonstrates how machine learning models based on $K(r)$-derived metrics can accurately estimate cluster size and intra-cluster density in simulated three dimensional (3D) point patterns containing spherical clusters of varying size; over 90\% of model estimates for cluster size and intra-cluster density fall within 11\% and 18\% error of the true values, respectively. These $K(r)$-based size and density estimates are then applied to an experimental APT reconstruction to characterize MgZn clusters in a 7000 series aluminum alloy. We find that the estimates are more accurate, consistent, and robust to user interaction than estimates from the popular maximum separation algorithm. Using $K(r)$ and machine learning to measure clustering is an accurate and repeatable way to quantify this important material attribute.

\end{abstract}

\keywords{Ripley's K-function, spatial statistics, aggregation, cluster detection, atom probe tomography, machine learning }

\clearpage
\setcounter{page}{1}
\pagebreak

\section{Introduction}
Understanding clustering (spatial aggregation) of certain species in materials is critical for studying structure-property relationships in materials science. Properties ranging from mechanical hardness to electronic transport can be impacted by molecular or atomic aggregation \cite{lee2011morphology, yang2003control, wu2001effects, chui2007molecular}. The size, shape, spacing, and other defining characteristics of such aggregation play a crucial roll in determining its impact on material properties; for example, 7000 series aluminum alloys containing zinc and magnesium, which are commonly used in aerospace applications, are known to have higher strength when alloying elements aggregate into clusters of appropriate size \cite{deschamps1998influence}. Because of their impact on material properties, characterizing cluster properties such as size, intra-cluster atomic concentration, background atomic concentration, and cluster spacing is of great utility to the materials science community. Atom Probe Tomography (APT), which produces sub-nanometer resolution 3-dimensional (3D) tomograms consisting of the $(x, y, z)$ positions and mass-to-charge-state ratios of each ion collected from a material sample, allows for detailed analysis of nanoscale morphology \cite{seidman2009atom}. In this work, we develop a procedure for accurate quantification of clusters in APT data using tools from point pattern spatial statistics. 

There are numerous existing cluster search algorithms designed to identify areas of aggregation in APT reconstructions \cite{marquis2010applications, felfer2015detecting, lefebvre2011application, stephenson2007new, zelenty2017detecting}. These methods generally rely on one or more initialization parameters from the user. For example, the maximum separation algorithm (MSA), one of the most commonly used cluster detection algorithms in the APT community, requires the initialization of two crucial parameters: $d_\text{max}$ (the maximum distance between two points in a cluster) and $N_\text{min}$ (the minimum number of points in a cluster) \cite{vaumousse2003procedure}. This algorithm's output is highly sensitive to the choice of these two parameters \cite{hyde2011sensitivity, dong2018atom}, and while there are general guidelines for this parameter selection \cite{stephenson2007new, vaumousse2003procedure}, the procedures are still subjective. This user-guided parameter selection is typical of other cluster detection algorithms as well \cite{marquis2010applications,lefebvre2011application}. The dependence of these algorithms on user judgment to select these critical parameters compromises the reproducibility of their results. 

Other methods have been developed to analyze clustering in APT reconstructions based on spatial statistics, using tools like the nearest neighbor distance distribution or radial distribution function \cite{philippe2013point, moody2008quantitative, philippe2010clustering, philippe2009clustering, marceau2011quantitative}. One of the most powerful spatial statistics tools for detection of clustering is Ripley's K-Function ($K(r)$), which summarizes spatial correlation in point patterns at a variety of length scales \cite{dixon2014r, baddeley2015spatial}. $K(r)$ has been used to estimate the size of circular point clusters in 2D point patterns \cite{lagache2013analysis, shivanandan2016characterizing, kiskowski2009use}, but there has been minimal work exploring how this extends to 3D (\textit{e.g.} for APT data analysis applications). 
$K(r)$-based cluster size estimates are independent of initialization parameters, and are thus largely independent of the user; this makes them attractive as a consistent tool for characterizing clusters.

In this work, we develop a general method using $K(r)$ to characterize clusters in 3D spatial point patterns, with a specific focus on application to APT data, where atoms or molecules can be represented by points in a pattern. We use summarizing metrics from $K(r)$ measured on simulated 3D point patterns with spherical clusters to train machine learning models to estimate cluster size, intra-cluster point concentration, and background point concentration; these models produce accurate estimates for point patterns with non-uniform cluster size, sporadic cluster spacing, and significant background point concentration. Applying the method to an APT reconstruction of a 7000 series AlMgZn alloy, we compare the resultant size and concentration estimates to similar estimates obtained using the MSA and find that the $K(r)$-based estimates offer a significant improvement over those from the MSA. This procedure is independent of user-defined initialization parameters, providing an accurate, reliable, and consistent method to characterize clustering in material systems.

\section{Background and Methods}

\subsection{Ripley's K-function}
Ripley's K-function measures correlations between point locations at different length scales in a point pattern. It is defined as the expected number of additional points that lie within a radius $r$ of a typical point in the pattern, normalized by the global intensity (\textit{i.e.} points per unit volume) of the pattern \cite{dixon2014r}. Given an observed point pattern, $K(r)$ is calculated as 
\begin{align}
    K(r) = \frac{1}{\lambda \, m}\sum_{i=1}^m \sum_{\substack{j = 1 \\ j \neq i}}^m I \{d_{ij} < r\} \, e_{ij}(r),
    \label{K}
\end{align}
where $m$ is the number of points in the pattern, $\lambda$ is the intensity of the point pattern ($m$/volume of pattern), $d_{ij}$ is the distance between points $i$ and $j$, $I$ is the indicator function (\textit{i.e.} $I \{d_{ij} < r\}$ is equal to one for $d_{ij} < r$ and equal to zero otherwise), and $e_{ij}(r)$ is an edge correction weight that adjusts for bias introduced by points that lie within a distance $r$ of the edge of the pattern \cite{baddeley2015spatial}. $K(r)$ is typically estimated at a finely spaced series of $r$ values over a length scale of interest.

In this study, we use $K(r)$ to analyze marked point patterns where each point is assigned a categorical label or ``mark" from one of two categories: $A$ or $B$. We denote the total number of points in a pattern as $m$, and the fraction of point marked type-$A$ as $\eta$, $0 < \eta < 1$. The set of all point positions in a pattern, disregarding their marks, is called the underlying point pattern (UPP). For the remainder of this work, we measure $K(r)$ on only the subset of type-$A$ points within the UPP; this allows us to study the spatial structure of type-$A$ points within the UPP. 

Random relabeling of the UPP allows for estimation of the expected $K(r)$ signal for randomly distributed marks within the pattern. Random relabeling is a process in which $K(r)$ is measured on a random subset of $m \eta$ points from the UPP. Repeating this process of subset selection and measurement many times creates an envelope showing where $K(r)$ measurements of randomly marked versions of the UPP are expected to fall; the median of this envelope is the expected $K(r)$ signal for a random distribution of type-$A$ points within the UPP. If the $K(r)$ measurement from the originally observed point pattern ($K_\text{A-obs}(r)$) deviates significantly from this envelope, one may reasonably conclude that the distribution of type-$A$ marks in the observed point pattern is not random \cite{baddeley2015spatial}. The goal of this work is to make inferences about the structure of type-$A$ marks within the UPP based on the behavior of this deviation from the random signal.

We use a transformed version of $K(r)$, which we call $T(r)$, to directly measure the deviation of the observed signal from the expected random signal:

\begin{align}
    T(r) = \sqrt{K_\text{A-obs}(r)} - \sqrt{K_{(N/2)}(r)} \,,
    \label{Kanom}
\end{align}

\noindent where $K_{(N/2)}(r)$ is the median of the $K(r)$ envelopes for $N$ random relabelings of the UPP; this subtraction serves to center $T(r)$ around zero. The square roots in equation (\ref{Kanom}) serve to stabilize the variance of $T(r)$ across all $r$ values \cite{baddeley2015spatial}. A more detailed discussion of the procedure used to calculate $T(r)$ is provided in Section \ref{SI_sec_kfn} of the SI.

\subsection{K-Function-Derived Metrics} \label{sec_kmets}
\begin{figure}[H]
    \centering
    \includegraphics[width = .5\linewidth]{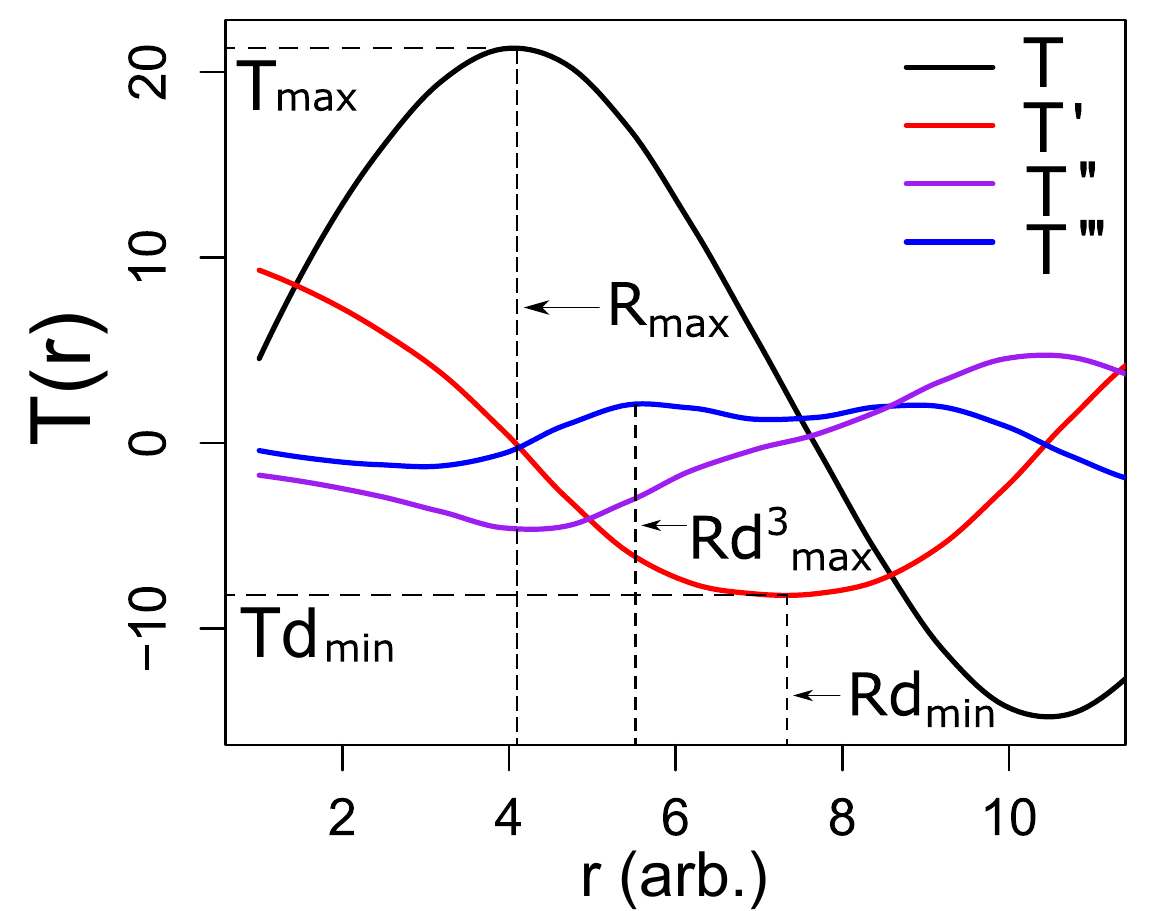}
    \caption{Example $T(r)$ measurement from a clustered point pattern (black line) along with its first three derivatives (colored lines). The metrics of interest are: $T_\text{max}$, height of the first maxima of $T(r)$; $R_\text{max}$, $r$ value at the first maxima of $T(r)$; $Td_\text{min}$, value of $T'(r)$ at its first minima;  $Rd_\text{min}$, $r$ value at the first minima of $T'(r)$; $Rd^3_\text{max}$, $r$ value at the first maxima of $T'''(r)$.}
    \label{kmetricsexample}
\end{figure}

Single metrics from $K(r)$, such as the radius of maximal aggregation \cite{shivanandan2016characterizing, kiskowski2009use} or functions of the derivatives of $K(r)$ \cite{kiskowski2009use}, have been used to estimate cluster size in 2D point patterns. These single-metric-based estimators showed promising results for ideal clustered patterns (uniform cluster size, regular cluster spacing, low background point concentration), but began to fail as non-idealities were introduced. To build a more robust model for cluster size, intra-cluster point concentration, and background point concentration, we instead use five separate metrics based on $T(r)$. Each of the five metrics is a function of $T(r)$ or its derivatives. The derivatives are calculated using the central difference method, and maxima and minima are located using rolling local polynomial fit smoothing. Table \ref{metrics} shows the name and description of each metric, and Figure \ref{kmetricsexample} shows an example $T(r)$ result from a type-$A$ clustered data set, its first three derivatives, and the location of each metric. $R_\text{max}$, $Rd_\text{min}$, and $Rd^3_\text{max}$ are correlated with cluster size \cite{shivanandan2016characterizing, kiskowski2009use}, while $T_\text{max}$ and $Td_\text{min}$ capture the strength of the $T(r)$ signal and how quickly it falls from its first peak - properties largely determined by intra-cluster and background point concentration.


\begin{table}[H]
\caption{Name and Description of $T(r)$ Metrics}
\label{metrics}
\centering
\begin{tabular}{ll}
\hline
\multicolumn{1}{c}{Name} & \multicolumn{1}{c}{Description}                \\ \hline
$T_\text{max}$           & The value of $T(r)$ at its first maximum.      \\
$R_\text{max}$           & The $r$ value where $T_\text{max}$ occurs.     \\
$Td_\text{min}$          & The value of $T'(r)$ at its first minimum   \\ 
$Rd_\text{min}$          & The $r$ value where $Td_\text{min}$ occurs.    \\
$Rd^3_\text{max}$        & The $r$ value at the first maximum of $T'''(r)$  \\ \hline
\end{tabular}
\end{table}

\subsection{Simulation of Clustered Point Patterns} \label{sec_cluster_sim}
To develop our cluster characterization method, we use simulated 3D point patterns with spherically clustered type-$A$ marks. The UPP in these simulations is taken to be the centers of 3D random close-packed (RCP) spheres; the diameter of these spheres defines the length unit for all simulated work in this study (denoted as ``arb."). This unit is scalable to any physical length, and is therefore somewhat arbitrary, but defined for consistency. The RCP structure emulates that of amorphous material systems; the points have a minimum nearest neighbor distance but do not follow any crystalline pattern at larger length scales. A previously developed algorithm \cite{desmond2009random} is used to generate realizations of RCP spheres within a $60 \times 60 \times 60$ arb. cubic volume. These UPPs have intensity of $\lambda_\text{sim} = 1.029$ points/arb.$^3$.

Each point in the UPP is then marked as either type-$A$ (cluster species) or type-$B$ (host species). A type-$A$ fraction of $\eta = 0.0511$ is used for all simulations in this study to match the measured $\eta$ for the APT reconstruction analyzed in Section \ref{sec_experimental}. Type-$A$ points are assigned within the UPP to create spherical clusters in the following way: the radius $R_c$ of each cluster within a pattern is a normally distributed random variable with mean $\mu_R$ and variance $\sigma_R^2$ (\textit{i.e.} $R_c \sim N(\mu_R, \sigma_R^2)$); cluster centroids are defined by the centers of a scaled-up RCP sphere pattern and are then shifted in a uniformly random direction by random distance $d \sim |N(0, \sigma_C^2)|$; the intra-cluster concentration of type-$A$ points (\textit{i.e.} type-$A$ points in a cluster/total points in the same cluster) is denoted $\rho_1$; the background concentration of type-$A$ points (\textit{i.e.} type-$A$ points outside of clusters/total points outside of clusters) is denoted $\rho_2$. The number of clusters within any simulated point pattern is determined by a combination of these five parameters. 

Note that the $\sigma_R$ and $\sigma_C$ parameters are reported alternatively as radius blur ($\beta = \sigma_R/\mu_R$) and position blur ($\xi = \sigma_C/\mu_\text{sep}$, where $\mu_\text{sep}$ is the average separation distance between cluster centers before they are shifted). For simulations in this study, we constrain $\beta < 0.5$ to avoid point patterns where many clusters are assigned a negative radius (when this occurs, the cluster is omitted completely from the simulation). Similarly, we constrain $\xi < 0.2$ to avoid having many clusters overlap after shifting their positions.

Simulations were performed in \texttt{R} mainly using the \texttt{spatstat} \cite{baddeley2015spatial} and \texttt{rapt} \cite{rapt2020} packages. Example simulation workflow scripts are available as Supplemental Information and on GitHub \cite{vignette2020}. A more detailed discussion of the cluster simulation process can be found in Section \ref{SI_sec_clustersim} of the SI.

\section{Simulation Results} \label{results}

\subsection{Weighted Radius} \label{sec_weighted_radius}
The mean cluster radius $\mu_R$ is difficult to estimate using $T(r)$ metrics when the clusters within a single sample vary in size (\textit{i.e.} $\beta \neq 0$) because larger clusters impact $K(r)$ (and therefore $T(r)$) disproportionately more than smaller clusters. Larger clusters (1) occupy a larger volume in the point pattern than the smaller clusters, and therefore contribute a ``large cluster" signal to $K(r)$; (2) contain more points than the smaller clusters, meaning that the ``large cluster" signals dominate $K(r)$, which is an average signal over every point in the pattern. The impact of this in point patterns with mean $\mu_R$ and non-uniform cluster sizes is that $K(r)$ will behave similarly to that from a point pattern with uniform cluster radii $\mu_R^*$, where $\mu_R^* > \mu_R$. This becomes an issue when trying to estimate $\mu_R$, as $T(r)$ metrics can not distinguish between these two situations.

Henceforth, we use the weighted radius ($R_w$), which is a function of the distribution of radii in a point pattern that gives each cluster the same weight that it receives in $K(r)$. This quantity can be generally written as

\begin{align}
    R_w  = \frac{\E\left[r \times \frac{4}{3} \pi r^3\right]}{\E\left[\frac{4}{3} \pi r^3 \right]},
    \label{rw_general}
\end{align}
where $\E\left[X\right]$ is the expected value of a random variable $X$. $R_w$ can be interpreted as the weighted average radius of clusters in the pattern, where the weight on each cluster's radius $r$ is its corresponding volume $\frac{4}{3} \pi r^3$. We use normally distributed radii ($R_c \sim N(\mu_R, \sigma_R^2)$), which simplifies equation (\ref{rw_general}) to the closed form expression
\begin{align}
    R_w = \frac{\mu_R^4 + 6 \mu_R^2 \sigma_R^2 + 3 \sigma_R^4}{\mu_R^3 + 3 \mu_R \sigma_R^2}.
    \label{rw_normal}
\end{align}

\noindent Because $K(r)$ directly measures $R_w$, which is a function of both $\mu_R$ and $\sigma_R$, it is impossible to individually extract $\mu_R$ or $\sigma_R$ given only output from $K(r)$. We therefore build models to estimate $R_w$ and leave the decoupling of $\mu_R$ and $\sigma_R$ for future work. 

\subsection{Estimating Cluster Properties} \label{sec_param_estimation}

In this section, we develop a model that uses the five $T(r)$ metrics introduced in Section \ref{sec_kmets} as predictors to estimate weighted radius ($R_w$), intra-cluster type-$A$ point concentration ($\rho_1$), and background type-$A$ point concentration ($\rho_2$) for a simulated clustered point pattern of interest. 

The $T(r)$ metrics were explored by sweeping through each simulation parameter ($\mu_R$, $\rho_1$, $\rho_2$, $\beta$, and $\xi$) individually, which revealed that the five $T(r)$ metrics relate to the simulation parameters nonlinearly and are significantly correlated, meaning that multicollinearity of the predictors would likely be an issue for a standard linear regression model. The full results of this exploration are provided in Section \ref{SI_sec_sweeps} of the SI.

To eliminate issues with correlation, we use principal component analysis to transform the $T(r)$ metrics into their standardized principal components (PCs) \cite{jolliffe2011principal}, and use these PCs as predictors in machine learning models for estimating $R_w$, $\rho_1$, and $\rho_2$. Machine learning allows us to capture the complex relationships between the $T(r)$-metrics and clustering parameters. To train these models, we generated 10,000 sets of random cluster parameters selected from uniform distributions over the ranges shown in Table \ref{parameterranges}. For each parameter combination, we simulated 10 clustered point patterns according to the procedure outlined in Section \ref{sec_cluster_sim}, resulting in a total of 100,000 simulated patterns. This method of repeating simulations at each parameter combination was designed to capture variation between simulations with the same input parameters. The expected random $K(r)$ signal was measured based on 50,000 random relabelings of the UPP, and was then used to calculate $T(r)$ and the five $T(r)$ metrics for each of the 100,000 simulated point patterns. We then calculated the five standardized PCs based on these metrics. A summary of the variance explained by these standardized PCs can be found in Section \ref{SI_sec_pca} of the SI. This set of 100,000 parameter-PC pairings were used to train the $R_w$, $\rho_1$, and $\rho_2$ models. 

\begin{table}[H]
\centering
\caption{Cluster Parameter Ranges}
\label{parameterranges}
\begin{tabular}{ll}
\hline
\multicolumn{1}{c}{Parameter}  & \multicolumn{1}{c}{Range} \\ \hline
Cluster Radius ($\mu_R$)             & {[}2, 6.5{]} arb.       \\
Intra-cluster Concentration ($\rho_1$)   & {[}0.2, 1{]}    \\
Background Concentration ($\rho_2$)       & {[}0, 0.03{]}    \\
Radius Blur ($\beta$)                       & {[}0, 0.5{]}  \\
Position Blur ($\xi$)                        & {[}0, 0.2{]}  \\ \hline
\end{tabular}
\end{table}

The ``no free lunch theorem" states that no single type of model will perform best for every data set \cite{murphy2012machine}. Therefore, we tested three separate popular regression models appropriate for this type of analysis: a generalized linear model (GLM), a random forest model (RF), and a Bayesian regularized neural network (BRNN) model. To compare these options, we used a testing data set of 25,000 simulated clustered point patterns, each with a unique set of parameters selected from uniform distributions over the same ranges shown in Table \ref{parameterranges}. 

We trained separate GLM, RF, and BRNN models for each parameter $R_w$, $\rho_1$, and $\rho_2$ in \texttt{R} using the \texttt{caret} package \cite{kuhn2008building} and 10-fold cross-validation. The best model for each parameter was selected as the one with the smallest root mean square error of prediction (RMSEP) for the testing data set. These RMSEP values are shown in Table \ref{model_RMSEP}; the BRNN model performed best for all three parameters. 

\begin{table}[H]
\centering
\caption{RMSEP for Different Models and Parameters}
\label{model_RMSEP}
\begin{tabular}{cllll}
\hline
\multicolumn{1}{l}{}   &       & $R_w$  & $\rho_1$ & $\rho_2$ \\ \hline
\multirow{3}{*}{RMSEP} & GLM   & 0.4678 & 0.0875   & 0.00549  \\
                       & BRNN  & 0.4142 & 0.0559   & 0.00372  \\
                       & RF    & 0.4272 & 0.0569   & 0.00381  \\ \hline
\end{tabular}
\end{table}

\begin{figure}[H]
    \centering
    \includegraphics[width=1\linewidth]{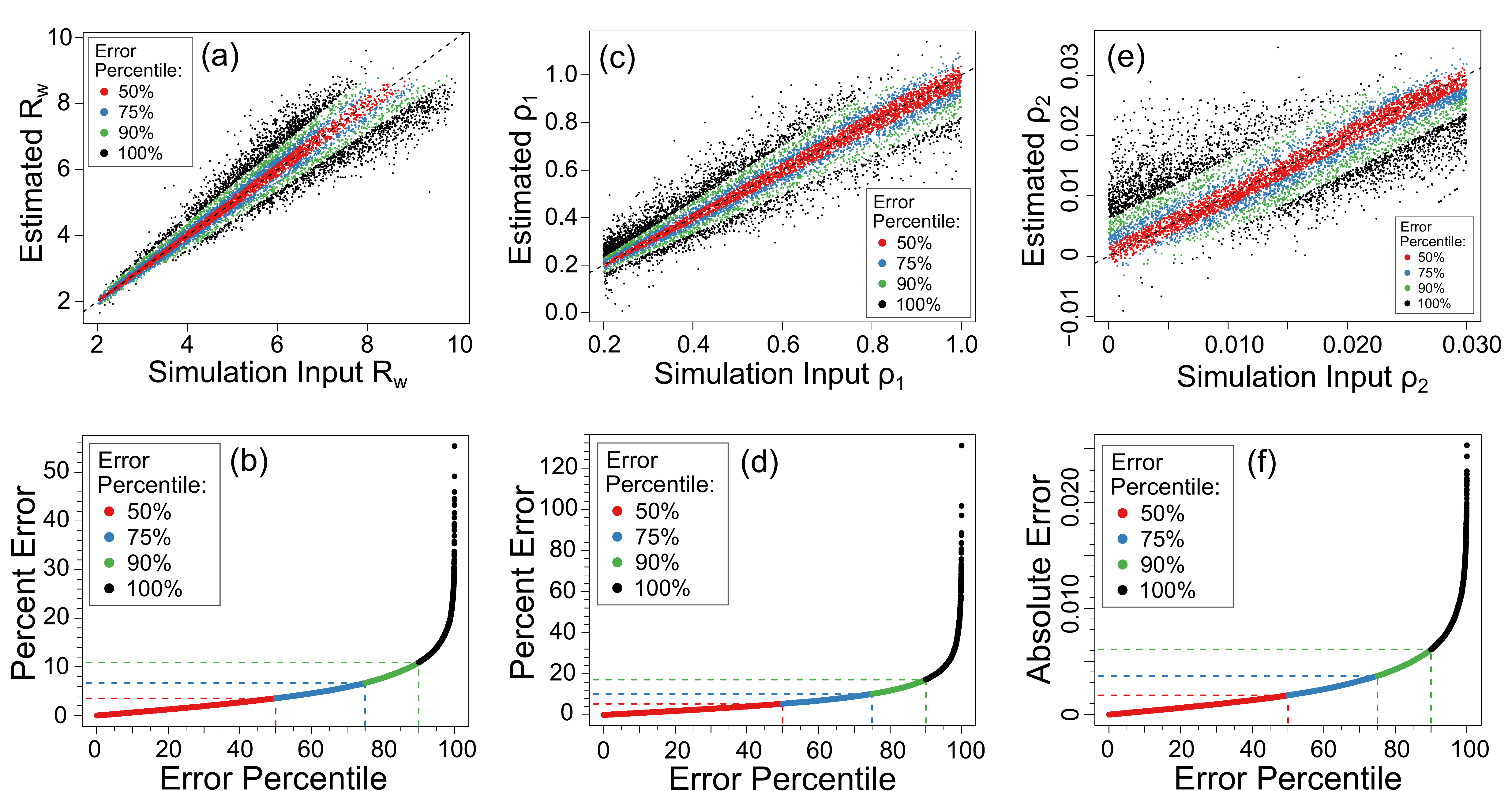}
    \caption{Results from the BRNN models of weighted radius ($R_w$), intra-cluster concentration ($\rho_1$), and background concentration ($\rho_2$); (a, c, e) Model estimates versus corresponding input simulated values for $R_w$, $\rho_1$, and $\rho_2$, respectively. Colors correspond to different error percentiles of the estimates (percent error for $R_w$ and $\rho_1$, absolute error for $\rho_2$); (b, d, f) Ordered error in testing data estimates for $R_w$, $\rho_1$, and $\rho_2$, respectively.}
    \label{mlmodels}
\end{figure}

Figure \ref{mlmodels} summarizes the performance of these BRNN models on the testing data set of 25,000 simulations. Figure \ref{mlmodels}(a) shows the true simulated values of $R_w$ versus the corresponding model estimates; the colors correspond to the percent error percentiles of these estimates (\textit{e.g.} the 50\% of estimates with lowest percent error are shown in red). Figure \ref{mlmodels}(b) shows the percent error of each estimate sorted in ascending order to easily determine what percent of model estimates fall below a certain percent error (\textit{e.g.} 50\% of estimates fall within 4\% error). Figures \ref{mlmodels}(c,d) show similar plots for the $\rho_1$ testing data estimates from the BRNN model. Figures \ref{mlmodels}(e,f) show similar plots for the $\rho_2$ testing data estimates from the BRNN model, except that absolute error is used in place of percent error because $\rho_2 \approx 0$ in many of the simulated point pattern leads to very large percent errors.

In an auxiliary simulation study, we used the same process described above to train models for simulated point patterns with uniform cluster radii ($\beta = 0$). These models have much higher predictive ability than the models presented above for non-uniform cluster size and are detailed in Section \ref{SI_sec_norb} of the SI. 

\subsection{Discussion of Simulation Results}

Table \ref{model_RMSEP} shows that the BRNN model has the smallest RMSEP on the testing data set for $R_w$, $\rho_1$, and $\rho_2$ estimates. Figure \ref{mlmodels} shows that for the $R_w$ model, 90\% of estimates from this BRNN model fall below 11\% error; for the $\rho_1$ model, 90\% of estimates fall below 18\% error; and for the $\rho_2$ model, 90\% of estimates fall within 0.007 (absolute) of the true parameter value. These results are promising for the $R_w$ and $\rho_1$ models, but less so for the $\rho_2$ model.



We have shown that precise and accurate characterization of cluster size and intra-cluster concentration using $K(r)$ is possible even in non-ideal cluster configurations (non-uniform cluster radii, irregular cluster spacing, and significant background point concentrations). It is noteworthy that the $\rho_1$ model was able to perform well even in these non-ideal situations, as $K(r)$ is not conventionally used as a measure of this quantity. The relatively poor performance of the $\rho_2$ model shows that the $T(r)$ metrics alone are weak predictors of the background point concentration of a clustered point pattern; this occurs because $K(r)$ contains more information about areas of high point concentration (\textit{i.e.} clusters) and less information about areas of low point concentration (\textit{i.e.} the space between clusters).

Cluster simulation studies are intrinsically limited by the parameters and simulation techniques, so we included a large range of clustering behaviors to cover many potential experimental outcomes; however, additional non-idealities could be explored, such as non-spherical clusters, non-uniform intra-cluster concentrations, or aberrations from the APT analysis and reconstruction processes. Furthermore, measures from other spatial statistics functions would likely supplement the $T(r)$-metrics to improve cluster parameter estimates. For example, the nearest-neighbor function $G(r)$ and empty-space function $F(r)$ measure small scale interactions in point patterns \cite{baddeley2015spatial}, which would be particularly useful for increasing the performance of the model for $\rho_2$.

\section{Application to Experimental Data} \label{sec_experimental}

In this section, we analyze an APT reconstruction of a 7000 series aluminum alloy with primary alloying constituents of magnesium and zinc (AlMgZn), a material commonly used in aerospace applications. Mg and Zn clustering of appropriate size and density in these alloys has been shown to increase material strength \cite{deschamps1998influence}. The reconstruction we analyze is from an APT sample run at 30 K in ultraviolet laser pulsing mode at 250 kHz. We apply our cluster characterization method to estimate the size and density of MgZn clusters present in the sample and compare these estimates to similar estimates obtained from the MSA over a variety of input parameter combinations. The precise experimental details behind this APT reconstruction are important to experimental work, but are outside the scope of this paper because our main goal is to address the efficacy of cluster property estimation within a reconstruction.


\subsection{MgZn Cluster Parameter Estimates}

After identifying and selecting (\textit{i.e.} ranging) only Al, Mg, and Zn atoms from the APT reconstruction, it contains 1,185,847 atoms in a 34.89 nm $\times$ 40.32 nm $\times$ 36.09 nm cuboid with atomic percentages (at\%) of 94.89\% Al, 2.78\% Zn, and 2.33\% Mg. The global intensity of the point pattern is $\lambda_\text{APT} = 23.36$ points/nm$^3$. 

It is typical in this type of alloy for Mg and Zn to cluster together \cite{sha2004early}, so for this analysis we treat Zn and Mg as a single aggregating species that makes up 5.11 at\% of the reconstruction. Our $K(r)$-based method could make independent estimates for each clustering species by training models using simulations containing multiple species, but for simplicity, we have chosen to look at the species in combination for this initial demonstration.

We only estimate $R_w$ and $\rho_1$ for this data set, as these models from Section \ref{sec_param_estimation} have the strongest performance. To apply these models to the AlMgZn alloy APT reconstruction, we assume that MgZn clusters in the sample are spherical, have normally distributed radii, and have uniform intra-cluster concentration. It is crucial that the spatial structure of the training data be consistent with that of the experimental data to apply these models accurately. The validity of these assumptions are addressed in Section \ref{SI_sec_assumptions} of the SI.

The experimental reconstruction and simulated training point patterns have significantly different global intensities ($\lambda_\text{APT} = 23.36$ points/nm$^3$, $\lambda_{sim} = 1.029$ points/arb.$^3$, respectively), which could lead to spurious estimates of $R_w$ and $\rho_1$, as the experimental estimates would be made in a parameter space that the model has not trained on. The experimental data was therefore scaled by a factor of $\alpha = \sqrt[3]{\lambda_\text{APT}/\lambda_\text{sim}} = 2.832$; this scaling preserves the features of the reconstruction but brings them into a length scale that our models have been trained on. The estimates of $R_w$ from this scaled reconstruction will be in simulation units (arb.); to get back to units of nm, the scaled $R_w$ estimate can be divided by $\alpha$. Intra-cluster concentration $\rho_1$ is independent of scaling, so the $\rho_1$ estimate for the scaled reconstruction is a valid $\rho_1$ estimate for the original reconstruction. 


We used the models trained in Section \ref{sec_param_estimation} to estimate $R_w$ and $\rho_1$ by measuring $T(r)$ on Mg and Zn atoms using 50,000 random relabelings of the measured UPP, extracting the five $T(r)$ metrics, and calculating the standardized metric PCs. These PCs were then used to obtain estimates of $R_w$ and $\rho_1$ for the scaled reconstruction. We used the training set to estimate uncertainty in these estimates by collecting the subsets of training data simulations that resulted in $R_w$ and $\rho_1$ estimates close to the estimates from the scaled reconstruction (within 0.2 arb. $R_w$ and 0.02 $\rho_1$), collecting the true $R_w$ and $\rho_1$ parameters from this subset of simulations, then using these true values to construct 90\% confidence intervals (CIs) for the true values of $R_w$ and $\rho_1$. The parameter estimates, along with the corresponding 90\% CIs are provided in Table \ref{apt_estimates}.

\begin{table}[H]
\centering
\caption{MgZn Cluster Parameter Estimates}
\label{apt_estimates}
\begin{tabular}{cccc}
\hline
Parameter & $R_w$ (arb.)       & $R_w$ (nm)         & $\rho_1$           \\ \hline
Estimate  & 5.192              & 1.834              & 0.212              \\
90\% CI   & {[}4.915, 5.700{]} & {[}1.736, 2.013{]} & {[}0.171, 0.255{]}  \\ \hline
\end{tabular}
\end{table}

\subsection{Comparison to Maximum Separation Algorithm}

It is informative to compare the estimates in Table \ref{apt_estimates} to estimates of the same parameters using the MSA, which is one of the most commonly used cluster identification tools in the APT community \cite{vaumousse2003procedure, hyde2011sensitivity, dong2018atom}. The MSA requires two user-defined input parameters: $d_\text{max}$, the maximum distance between two points in a cluster, and $N_\text{min}$, the minimum number of points in a cluster \cite{hyde2011sensitivity}. There are two other parameters used for enveloping background points into clusters: $L$ and $E$; but these parameters do not impact algorithm output significantly and are usually both set equal to $d_\text{max}$, which is what we do here \cite{hyde2011sensitivity}. To perform the MSA in this study, we used the \texttt{msa()} function within the \texttt{rapt} package \cite{rapt2020}.

The MSA only identifies which points reside in which clusters; to estimate $\rho_1$ and cluster size, further analysis is required. To estimate $\rho_1$ for each cluster identified by the MSA, we use the estimator:
\begin{align}
    \hat{\rho_1} = \frac{\text{number of type-$A$ points in cluster}}{\text{total number of points in cluster}}.
    \label{rho_estimate}
\end{align}
To estimate cluster radius for each cluster identified by the MSA, we use the estimator detailed in \cite{kolli2007comparison}:
\begin{align}
    \hat{R_c} = \sqrt{\frac{5}{3} \sum_{i = 1}^n r_i^2},
    \label{rc_estimate}
\end{align}
where $n$ is the number of type-$A$ points in the cluster and $r_i$ is the distance between point $i$ and the cluster center of mass. 

\begin{figure}[H]
    \centering
    \includegraphics[width=1\linewidth]{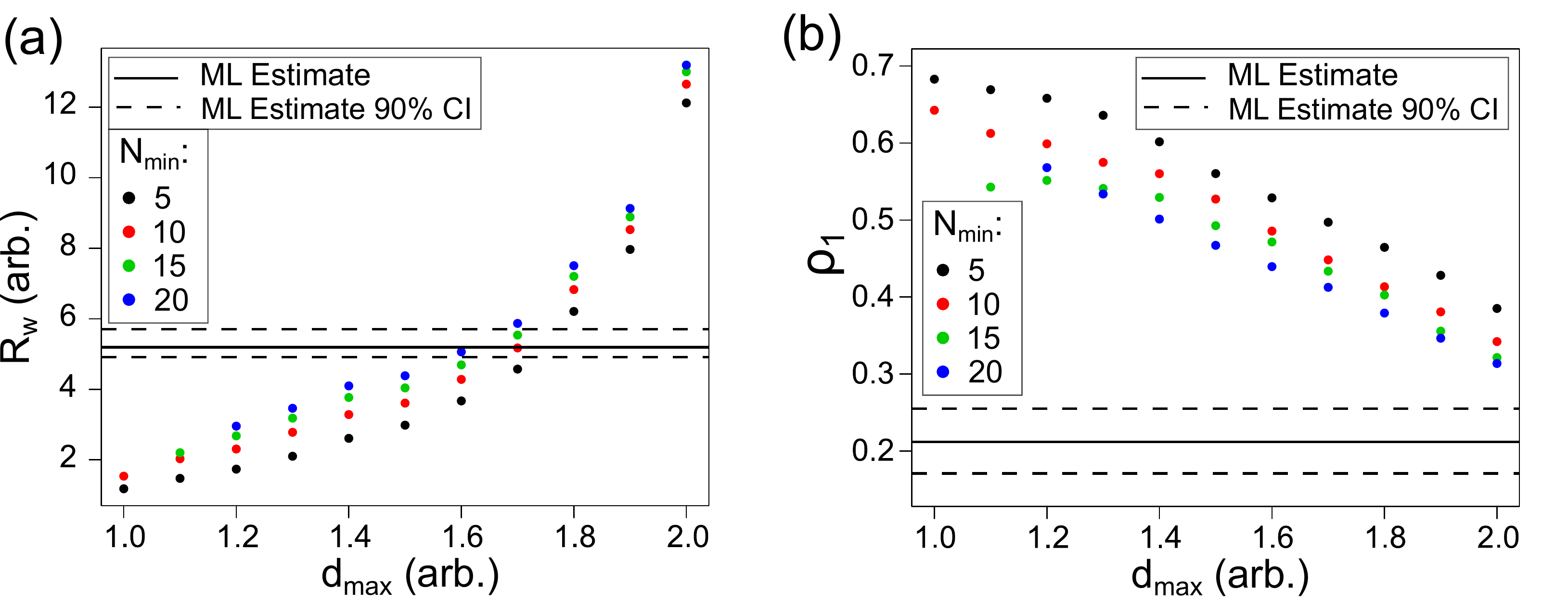}
    \caption{Maximum separation algorithm estimates for (a) weighted radius $R_w$ and (b) mean intra-cluster concentration $\rho_1$ for clusters in the scaled AlMgZn alloy APT reconstruction, shown as a function of $d_\text{max}$ and $N_\text{min}$. Solid black lines are estimates from $K(r)$-based machine learning (ML) model. Dashed black lines are the 90\% CI for the ML model estimates.}
    \label{msa_rw_den}
\end{figure}

We ran the MSA on the scaled version of the AlMgZn reconstruction using a range of $d_\text{max}$ and $N_\text{min}$ parameters, as recommended by the analysis of Vaumousse, et al. \cite{vaumousse2003procedure}. For each version of the MSA, the intra-cluster concentration $\rho_1$ and mean cluster radius $R_c$ and of all identified clusters in the sample were estimated using equations (\ref{rho_estimate}) and (\ref{rc_estimate}), respectively. A figure showing the mean $R_c$ estimate as a function of $d_\text{max}$ and $N_\text{min}$ is provided in Section \ref{SI_sec_msa} of the SI. Using the estimated $R_c$ for each MSA parameter combination, the weighted radius of the reconstruction $R_w$ was then calculated using equation (\ref{rw_general}). Figure \ref{msa_rw_den} shows these MSA estimates for $R_w$ and mean $\rho_1$ of the reconstruction as a function of $d_\text{max}$ and $N_\text{min}$ along with the $K(r)$-based estimates and 90\% CIs from Table \ref{apt_estimates}.

\begin{figure}[H]
    \centering
    \includegraphics[width=1\linewidth]{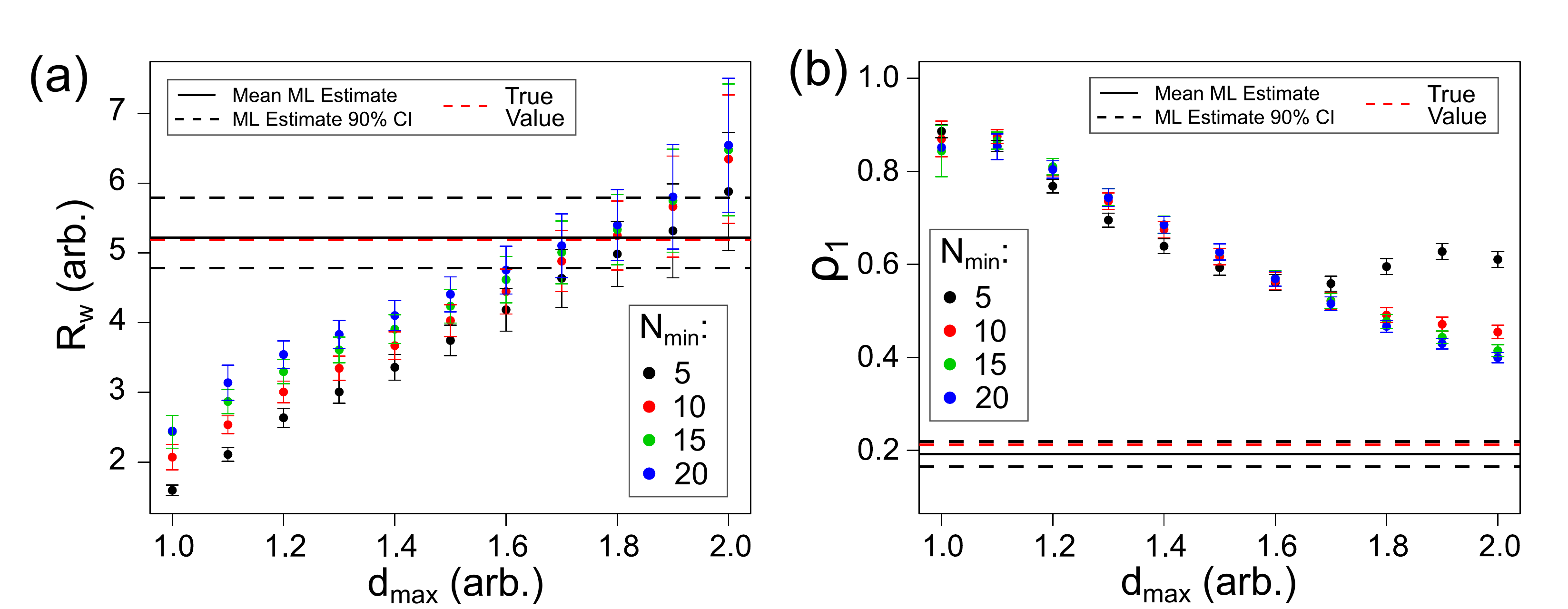}
    \caption{Maximum separation algorithm estimates for (a) weighted radius $R_w$ and (b) mean intra-cluster concentration $\rho_1$ for clusters in simulated point patterns with known $R_w = 5.192$, $\rho_1 = 0.212$, $\rho_2 = 0.03$, $\xi = 0$. True values are shown as red dashed lines. $K(r)$-based machine learning model estimates are shown as black lines with corresponding 90\% CIs shown as black dashed lines.}
    \label{msa_on_sims}
\end{figure}

To compare the behavior of these $K(r)$-based estimates with those from the MSA, we use results from simulated clustered data where the true values of $R_w$ and $\rho_1$ are known. We simulated 500 clustered point patterns with cluster species fraction $\eta = 0.0511$ and parameters $\mu_R = 4.5$, $\beta = 0.238$ ($R_w = 5.192$), $\rho_1 = 0.212$, $\rho_2 = 0.03$, and $\xi = 0$, and estimated $R_w$ and $\rho_1$ for each of these simulations using both our $K(r)$-based machine learning estimates and estimates from the MSA over the same range of $d_\text{max}$ and $N_\text{min}$ parameters used in Figure \ref{msa_rw_den}. Note that the simulated cluster parameters match those measured from the reconstruction (Table \ref{apt_estimates}). Figure \ref{msa_on_sims} shows the results from this comparison; the true values for $R_w$ and $\rho_1$ are shown as red dashed lines, the machine learning (ML) estimates of mean and 90\% CIs are shown as black lines, and the MSA estimates are shown as colored dots.

\subsection{Discussion of Experimental Results}
Figure \ref{msa_rw_den} shows that cluster property estimates of MgZn clustering in the reconstruction based on the MSA are highly dependent on $d_\text{max}$ and $N_\text{min}$, and are especially sensitive to $d_\text{max}$. Over the input parameter ranges swept, $R_w$ estimates ranged between 1 and 13 arb., while $\rho_1$ estimates ranged between 0.7 and 0.3. The $K(r)$-based $R_w$ estimate shown in Table \ref{apt_estimates} falls in line with the MSA $R_w$ estimates for certain parameter combinations ($d_\text{max} \approx 1.7$, $N_\text{min} \approx 10$), but the $K(r)$-based $\rho_1$ estimate shown in Table \ref{apt_estimates} does not match the MSA estimates for any $d_\text{max}$ and $N_\text{min}$ combination tried. 

When tested on simulated data with similar morphology to the reconstruction, the $K(r)$-based machine learning estimates match the true values of $R_w$ and $\rho_1$ closely with small variance. Similar to the experimental data, the MSA estimates for the simulated data vary significantly based on input parameter selection, only agree with the true value of $R_w$ for a select few parameter combinations, and fail to agree with the true value of $\rho_1$ for any input parameter combination tried. It is important to note that there was no $d_\text{max}$/$N_\text{min}$ combination that provided simultaneously accurate estimation of both $R_w$ and $\rho_1$, meaning that there is a fundamental disconnect between the clustering morphology identified by the MSA and the true clustering morphology of the system. Figures \ref{msa_rw_den} and \ref{msa_on_sims} show the similarities in behavior between the $K(r)$-based and MSA estimation methods performed on the experimental and simulated data. The accuracy and precision of the $K(r)$-based estimates for $R_w$ and $\rho_1$ for the simulated data paired with the agreement in estimate behavior between simulation and experimental data leads to the conclusion that the $K(r)$-based estimates agree closely with the true values of $R_w$ and $\rho_1$ in the reconstruction and should be trusted above the estimates calculated from the MSA. 

The consistent underestimation of the true value of $\rho_1$ in the simulated data (and likely the experimental data) by the MSA estimator is due to a high concentration of background cluster type points (\textit{i.e.} a large $\rho_2$ value). These background type-$A$ points are randomly distributed through the point pattern, so if their concentration is high, some are bound to fall within $d_\text{max}$ of the edge of a cluster. The MSA then includes these points as part of the identified cluster, resulting in small ``arms" of background type-$A$ points extruding from the true cluster spheres. These arms contain few type-$B$ points, so the clusters have a higher proportion of type-$A$ points, resulting in an upward bias of the $\rho_1$ estimate. If $d_\text{max}$ is increased, these arms smooth out and more type-$B$ points are enveloped into the cluster, resulting in lower estimates of $\rho_1$. This explanation is consistent with Figure \ref{msa_on_sims}(b), which shows the MSA $\rho_1$ estimates consistently overestimate the true value of $\rho_1$, but decrease towards the true value as $d_\text{max}$ is increased. Users of the MSA should be especially careful in point patterns with high background concentration, which are common in APT data.

A drawback to simulation-trained machine learning parameter estimation methods is that they only work well when the training data is consistent with the experimental data being analyzed. In this specific case, this idea corresponds to the cluster simulations agreeing with the morphology of a material. In contrast, the MSA can return clustering information for any morphology, which can sometimes be useful. However, Figures \ref{msa_rw_den} and \ref{msa_on_sims} clearly show that the estimates derived from the MSA are highly variable based on user-input parameters, while $K(r)$-based estimates provide significantly greater accuracy and confidence. By expanding simulations to include a more diverse range of clustering behaviors, it should be possible to obtain machine learning models that outperform the MSA on most types of morphologies. The use of point pattern measurement tools other than $K(r)$ should further increase the accuracy and reliability of these models.

\section{Conclusion}
Machine learning models based on metrics derived from Ripley's K-function can estimate cluster properties including size and intra-cluster concentration with high accuracy in simulated 3D data sets with non-ideal conditions. Over 90\% of $K(r)$-based estimates for $R_w$ and $\rho_1$ fell within 11\% and 18\% error of the true values, respectively. We applied this estimation method to an APT reconstruction of a 7000 series Al alloy to estimate $R_w$ and $\rho_1$ for MgZn clusters and compared these estimates to those from the popular MSA method, showing that estimates from the MSA strongly depend on user input parameters, while the $K(r)$-based estimates are more consistent. Using simulated data similar in morphology to the reconstruction, we then showed that the $K(r)$-based estimates are likely more accurate and consistent with the true values of $R_w$ and $\rho_1$ than those from the MSA method. These demonstrations show how this procedure for quantifying clustering in 3D point patterns can measure material morphologies more consistently and accurately than existing methods. Understanding the connection between a material's morphology and its properties is crucial for optimizing current materials and developing new ones; the procedures developed in this work can be used to further develop accurate structure-property relationships. The results here justify further investigation into the use of machine learning and spatial statistics to characterize tomographic data. 

\section{Acknowledgements}
The authors would like to thank Stephan Gerstl and the Scientific Center for Optical and Electron Microscopy at ETH Z\"urich for supplying the AlMgZn APT reconstruction analyzed in this work. The authors would additionally like to thank Matthew B. Jaskot and Paul Niyonkuru for their manuscript feedback and suggestions.

\section{Author Contributions}
GBV - Method development, simulation development and execution, analysis of experimental data, interpretation of results, primary manuscript writing and editing. 

\noindent APP - Method development, manuscript editing.

\noindent JDZ - Method development, interpretation of results, manuscript editing.

\section{Funding Sources}
Work by GBV, APP, and JDZ was supported by the U.S. Department of Energy, Office of Science, Basic Energy Sciences under Award DE-SC0018021 and by funding from Universal Display Corporation. Funding sources had no involvement in this work.

\section{Competing Interests}
The authors have no competing interests to declare.

\newpage

\bibliography{bibliography}
\bibliographystyle{unsrt}

\makeatletter\@input{xxMain.tex}\makeatother
\end{document}


\maketitle
\vspace{-3em}

\section{Ripley's K-function} \label{SI_sec_kfn}
For an observed point pattern where a fraction $\eta$ of points are marked as type-$A$ and the remaining fraction $(1-\eta)$ of points are marked as type-$B$, the expected signal for a random distribution of type-$A$ points within the UPP can be determined as follows. $N$ random and independent subsets of points from the UPP are selected, each containing a fraction $\eta$ of the total number of points in the UPP. These subsets are selected such that each point in the UPP has an equal probability of being selected. $K(r)$ is measured on each of the $N$ resulting subsets at a series of $p$ finely spaced $r$ values $r_1, \ldots, r_p$. The results can be stored in a $p \times N$ matrix $\mathbf{K}$, where the $i^{th}$ column of $\mathbf{K}$ contains the $K(r)$ measurement for the $i^{th}$ subset of the UPP. We then create a new $p \times N$ matrix $\mathbf{K}^*$, where the $j^{th}$ row contains the $j^{th}$ row of $\mathbf{K}$ sorted in ascending order. 

Let $\mathbf{k}^*_j$ represent the $j^{th}$ column of $\mathbf{K}^*$. Then $\mathbf{k}^*_{N(\alpha/2)}$ and $\mathbf{k}^*_{N(1 - \alpha/2)}$ contain the lower and upper bounds, respectively, of the $\alpha$-level acceptance interval (AI) envelope for random type-$A$ mark placement within the UPP \cite{baddeley2015spatial_SI}. 

 The measured $K(r)$ from type-$A$ points in the original observed pattern ($K_\text{A-obs}(r)$ for $r = r_1, \ldots, r_p$) is then plotted on top of the $\alpha$ level AI envelopes; if $K_\text{A-obs}(r)$ falls outside of the envelopes, there is evidence to reject the null hypothesis that type-$A$ points are randomly assigned within the UPP at the $\alpha$ level. More specifically, if $K_\text{A-obs}(r)$ deviates above or below the AI envelopes, then there is evidence of clustering or inhibition, respectively \cite{baddeley2015spatial_SI}. 

 Our transformed version of $K_\text{A-obs}(r)$, denoted $T(r)$, for each $r = r_1, \ldots, r_p$ can be calculated as
\vspace{-0.5em}
\begin{align}
    T(r_i) = \sqrt{K_\text{A-obs}(r_i)} - \sqrt{k^*_{i,N/2}}
\end{align}

\vspace{-0.5em}
\noindent where $k^*_{i,j}$ is the entry in the $i^{th}$ row and $j^{th}$ column of $\mathbf{K}^*$. Note that the $1 \times p$ vector $\mathbf{k^*_{N/2}}$ contains the expected $K(r)$ signal for randomly assigned type-$A$ points, as mentioned in the main body of the paper. Also note that the expected $T(r)$ signal for randomly assigned type-$A$ points is exactly $T(r) = 0$, hence the advantage of using such a transformation. 

\begin{figure}[H]
    \centering
    \includegraphics[width = 1\linewidth]{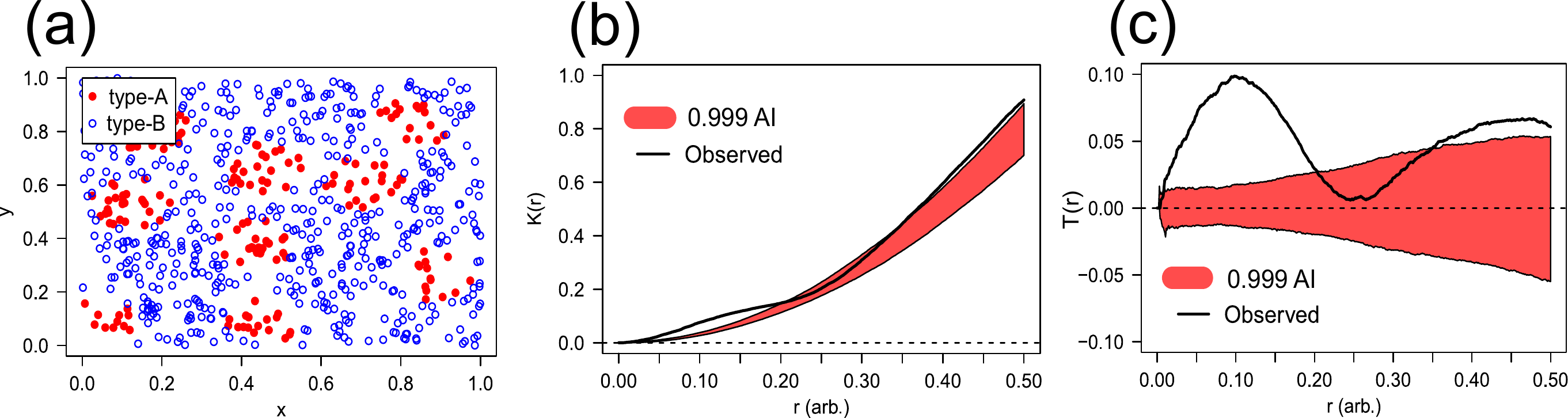}
    \caption{(a) An example marked point pattern with type-$A$ and type-$B$ points, where type-$A$ points are clustered. (b) The $\alpha = 0.999$ level $K(r)$ random relabeling AI envelopes from the UPP of the point pattern shown in (a) with $K_\text{A-obs}(r)$ superimposed. (c) The transformed $\alpha = 0.999$ level random relabeling AI envelopes from the UPP of the point pattern shown in (a) with $T(r)$ superimposed.}
    \label{kexample}
\end{figure}

Figure \ref{kexample} shows an example of (a) a 2D marked point pattern with points of type $A$ and $B$ where type-$A$ points are assigned in a clustered structure, (b) the $\alpha = 0.999$ level $K(r)$ random relabeling AI envelope with $K_\text{A-obs}(r)$ superimposed, and (c) the transformed $\alpha = 0.999$ level random relabeling AI envelope with $T(r)$ superimposed. Figures \ref{kexample}(b) and \ref{kexample}(c) display the same information, but the transformation applied in \ref{kexample}(c) highlights the important deviations from the expected random signal.

\section{Cluster Simulation Algorithm} \label{SI_sec_clustersim}

\begin{figure}[H]
    \centering
    \includegraphics[width = .5\linewidth]{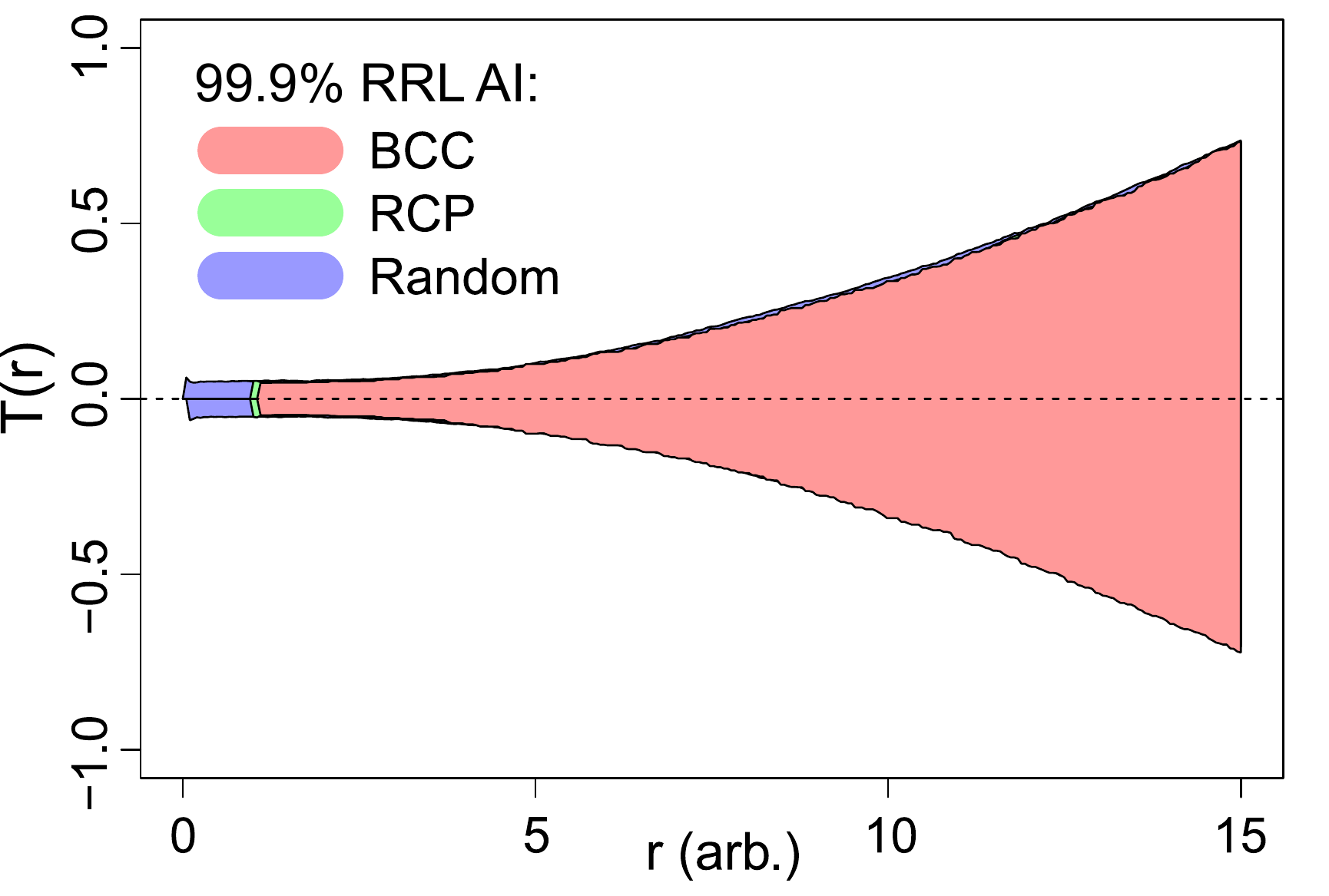}
    \caption{Comparison of transformed level 0.999 AI envelopes for $\eta = 0.0511$ random relabeling of body centered cubic (BCC), random close packed (RCP), and random UPPs. The only significant deviation between the envelopes is the location of their turn-on radius, which is determined by the minimum separation distance between points in the pattern.}
    \label{background_compare}
\end{figure}

 As noted in the main body of this text, 3D random close packed (RCP) sphere centers are used as the UPP for these simulations. It is important to note that the UPP has little impact on $T(r)$, so choice of UPP has little impact on the results of the study. This point is illustrated in Figure \ref{background_compare}, which shows the transformed level 0.999 AI envelopes for random relabeling ($\eta = 0.0511$) of body centered cubic (BCC), RCP, and random (Poisson point process) UPPs. The shape of the $T(r)$ envelopes is unchanged for each UPP configuration, minus their ``turn-on" radii, which is the minimum separation distance between points in the pattern. An RCP UPP was used in this work only because of its similarity to the structure of molecules or atoms in amorphous materials, but any UPP would return equivalent results at clustering size scales $r \gtrsim 2 \textrm{ arb.}$.

 RCP structures for the UPP in this work are generated using a previously developed algorithm that uses two different sphere sizes to minimize crystalline domains \cite{desmond2009random_SI}. We used 75\% 1 arb. diameter and 25\% 1.2 arb. diameter spheres, which resulted in only 6.5\% of points residing in crystalline regions according to a spherical harmonic-based test for crystallinity \cite{kurita2010experimental_SI}. The RCP generation algorithm was configured to return point patterns in a $20 \times 20 \times 20$ arb. cube with periodic boundaries, so that they can be tiled together to create cuboid UPPs of any size. These simulated UPP point patterns have intensity $\lambda_\text{sim} = 1.029$ points/arb.$^3$.

\begin{figure}[H]
    \centering
    \includegraphics[width = .85\linewidth]{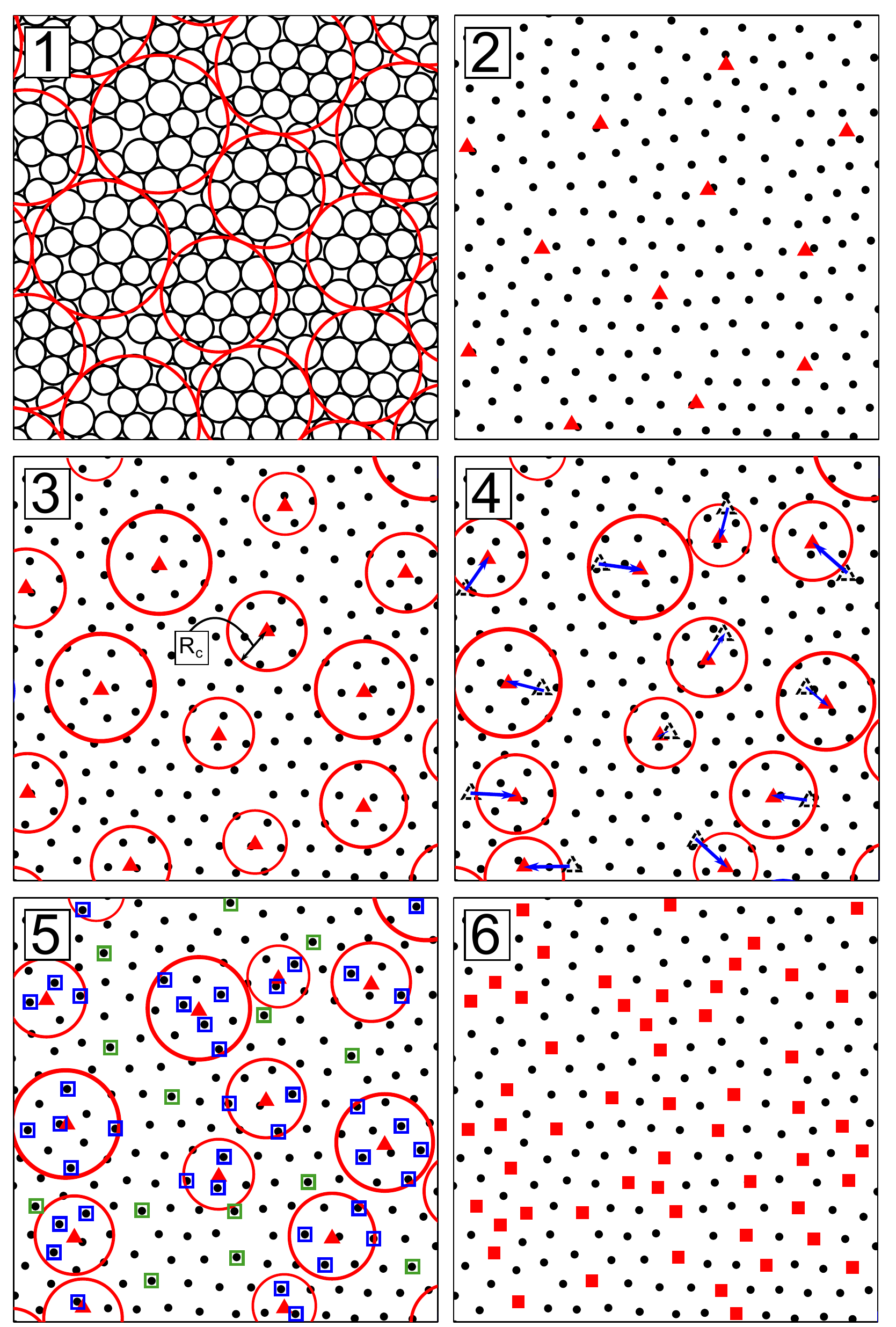}
    \caption{Step by step procedure for the cluster simulation algorithm in 2D.}
    \label{clustersim}
\end{figure}

Figure \ref{clustersim} shows an example in two dimensions of the cluster simulation process used in this work. The steps are: (1) Two random close packed circle patterns are overlaid. One is scaled with respect to the other so that that there will be approximately the correct number of type-$A$ points in the pattern at the end of the algorithm (scaling factor is a function of the input parameters). (2) The centers of the black circles are taken as the UPP, while the centers of the red circles are taken as the cluster centroids. (3) Normally distributed radii, R$_\text{c} \sim  N(\mu_R, \sigma_R^2)$, are assigned to each of the cluster centroids. These radii are shown in Panel 3 as red circles around the cluster centroids. (4) Cluster centroids are each shifted in a uniformly random direction by a normally distributed distance $d \sim |N(0, \sigma_C^2)|$. After this process is applied, any overlapping clusters are shifted away from one another until they no longer overlap. (5) A fraction $\rho_1$ of the UPP points within radius $R_c$ of each cluster centroid are marked as type-$A$, as shown by the blue squares in Panel 5. Similarly, a fraction $\rho_2$ of the UPP points in the background matrix are marked as type-$A$, as shown by the green squares in Panel 5. (6) The cluster centroids are removed from the pattern, leaving the final marked point pattern with spherical type-$A$ clusters. This procedure was extended into three dimensions using random close packed spheres in a 3D volume. Cluster simulation has been packaged into the \texttt{clustersim()} function of the \texttt{rapt} \texttt{R} package \cite{rapt2020_SI}.

\subsection{Simulation Volume Impact on T(r)}
The number of points in a point pattern has significant impact on the variability of $T(r)$ in samples with the same global properties (\textit{i.e.} simulations with the same input parameters but different random seeds). The shape of the sample volume also impacts the variability due to edge effects. Minimizing variability in $T(r)$ decreases the variability of $T(r)$ metrics, which leads to more accurate estimations of cluster properties. The downside of increasing sample size is that the computational cost of calculating $T(r)$ increases significantly with each added point. For a pattern with $n$ total type-$A$ points, computational complexity for calculating $K(r)$ is $O(n^2)$. 

To select a sample size that balances variance and computation time for our simulations, cuboid clustered point patterns with sizes ranging from 15$\times$15$\times$15 arb. to 60$\times$60$\times$60 arb. were simulated while tracking (1) the width of the $\alpha = 0.999$ $T(r)$ AI envelopes from $N = 50,000$ random relabelings of $\eta = 0.0511$ type-$A$ points in our RCP UPP, and (2) the standard deviation of each $T(r)$ metric from 500 simulated clustered patterns with parameters $\mu_R = 3$, $\rho_1 = 0.50$, $\rho_2 = 0$, $\beta = 0$, and $\xi = 0$.

\begin{figure}[H]
    \centering
    \includegraphics[width = .6\linewidth]{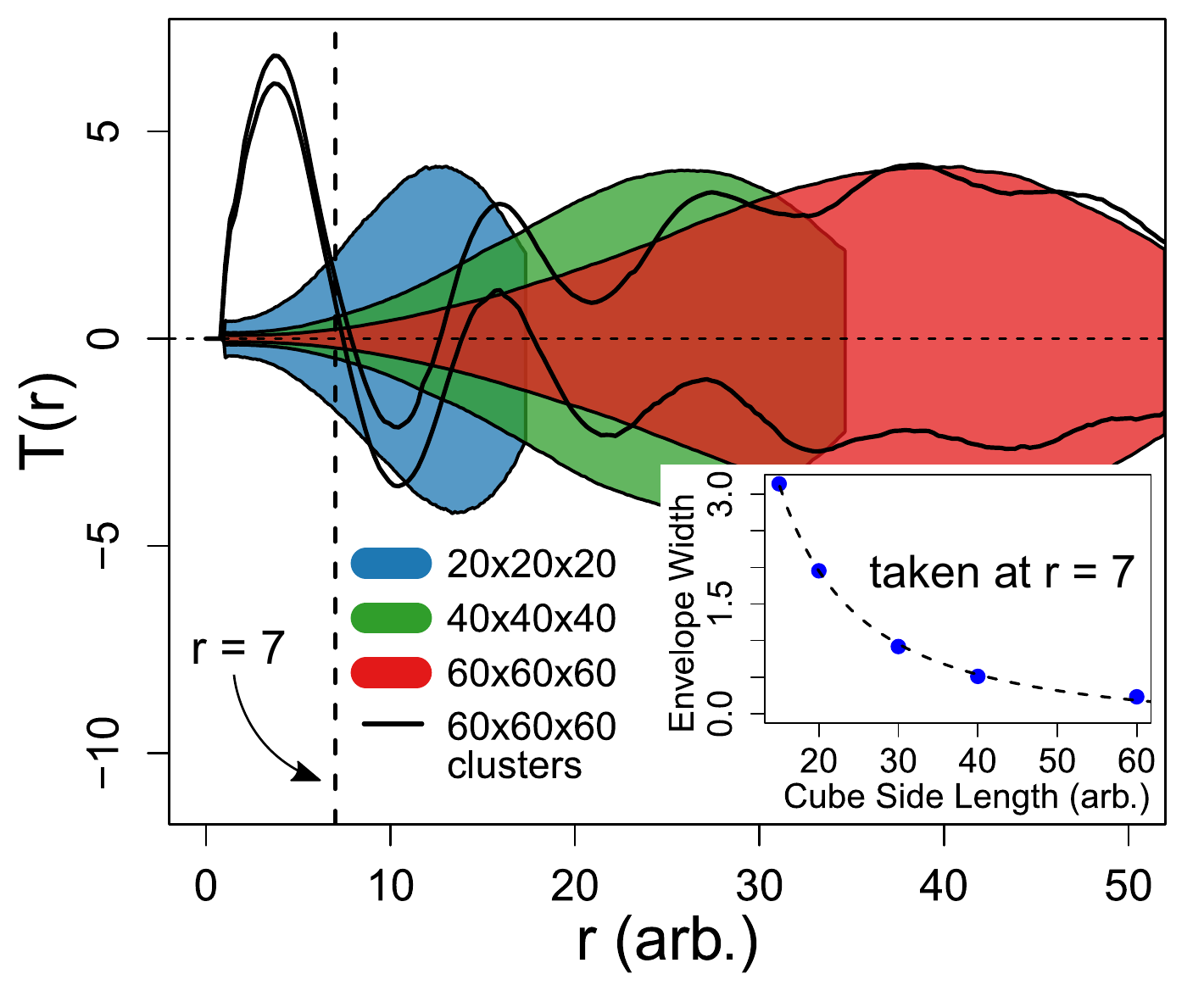}
     \caption{Comparing RRL AI envelopes for different sample volumes. The colored envelopes are the level 0.999 random relabeling AIs, and the black lines are the level 0.90 Monte Carlo cluster AI (for 60$\times$60$\times$60 arb. only), created by simulating 500 clustered patterns with $\mu_R = 3$, $\rho_1 = 0.50$, $\rho_2 = 0$, $\beta = 0$, $\xi = 0$, and measuring $T(r)$ on each of them. The bottom right plot inset shows the width of the RRL envelopes at $r = 7$ decaying as a function of cube side length. A (\textit{side length})$^{-3/2}$ fit is shown as a guide to the eye.}
    \label{envelopes}
\end{figure}

Figure \ref{envelopes} shows how the width of the $T(r)$ AI envelope shrinks as a function of simulation volume. The black lines superimposed over the colored AI envelopes is a level 0.90 AI envelope of $T(r)$ for the 500 simulated clustered data sets. The plot inset shows how the random relabeling $T(r)$ AI envelope width drops following approximately (\textit{side length})$^{-3/2}$. This is due to the fact that variance in $T(r)$ is roughly $\propto \,n^{-1/2}$, where $n$ is number of type-$A$ points in the pattern. For cubic volumes, $n \propto$ (\textit{side length})$^3$.

\begin{figure}[H]
    \centering
    \includegraphics[width = .6\linewidth]{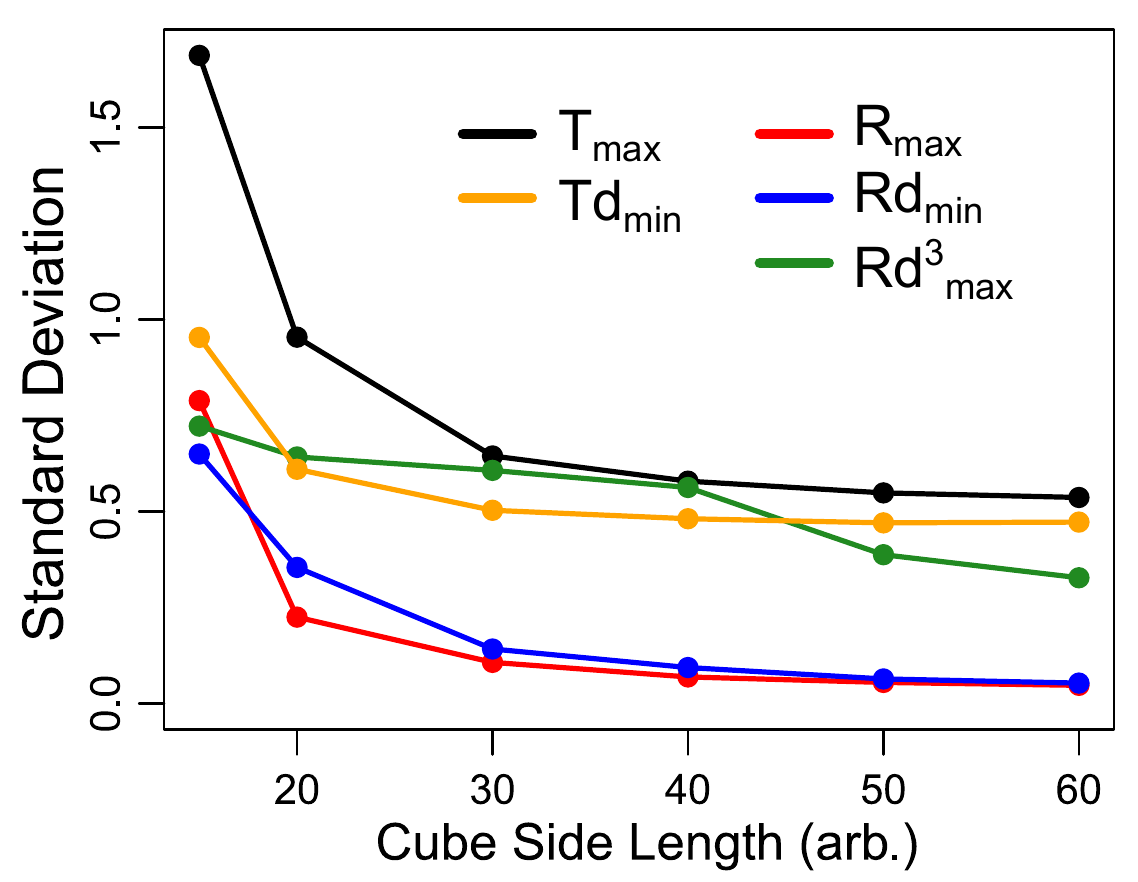}
     \caption{Standard deviation of the five $T(r)$ metrics from 500 cluster simulations with $\mu_R = 3$, $\rho_1 = 0.50$, $\rho_2 = 0$, $\beta = 0$, and $\xi = 0$ plotted versus cubic simulation volume side length.}
    \label{metric_width}
\end{figure}

Figure \ref{metric_width} shows how each $T(r)$ metric standard deviation for simulated clusters shrinks as a function of box size. The smaller the width, the less error in models based on these metrics. The AI width and $T(r)$ metric standard deviations see only minor improvements in patterns above 60$\times$60$\times$60 arb. size, but the computation time grows significantly; therefore we chose to perform simulations on this volume.  

\section{Sweeping Individual Cluster Parameters} \label{SI_sec_sweeps}

\begin{figure}[H]
\centering
\includegraphics[width = 1\linewidth]{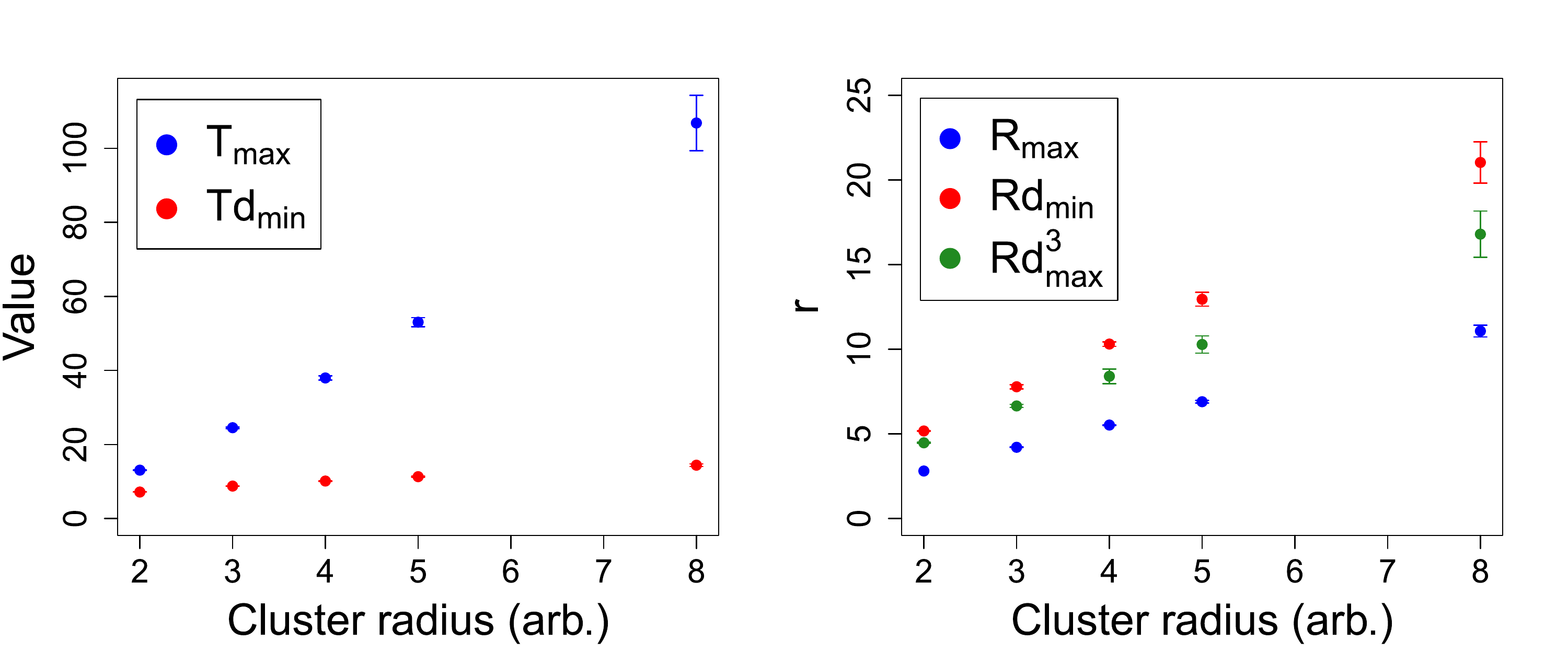}
\caption{$T(r)$ metrics measured on simulated clustered data sets while sweeping through a range of cluster radius values. Other simulated cluster parameters are: $\rho_1 = 1$, $\rho_2 = 0$, $\beta = 0$, $\xi = 0$.}
\label{series_radius}
\end{figure}

\begin{figure}[H]
\centering
\includegraphics[width = 1\linewidth]{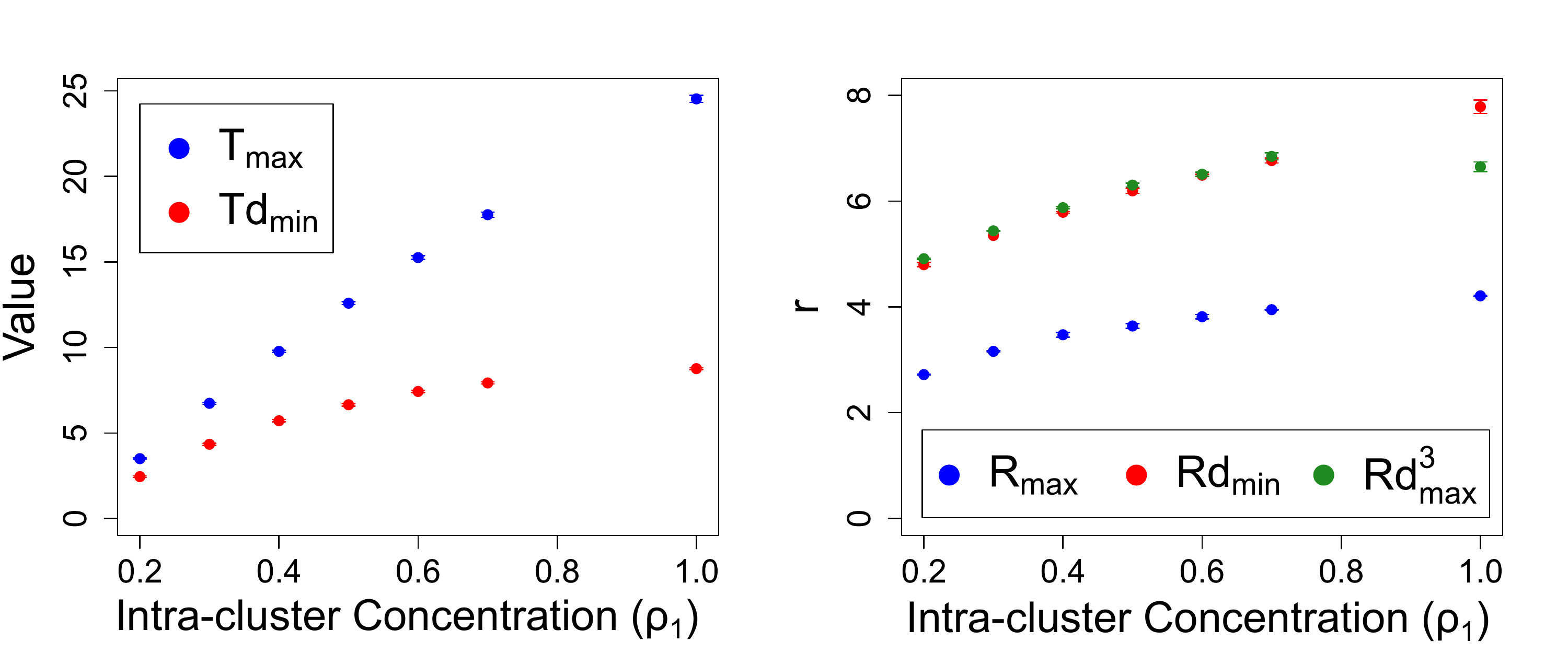}
\caption{$T(r)$ metrics measured on simulated clustered data sets while sweeping through a range of intra-cluster concentration values. Other simulated cluster parameters are: $\mu_R = 3$ arb., $\rho_2 = 0$, $\beta = 0$, $\xi = 0$.}
\label{series_rho1}
\end{figure}

\begin{figure}[H]
\centering
\includegraphics[width = 1\linewidth]{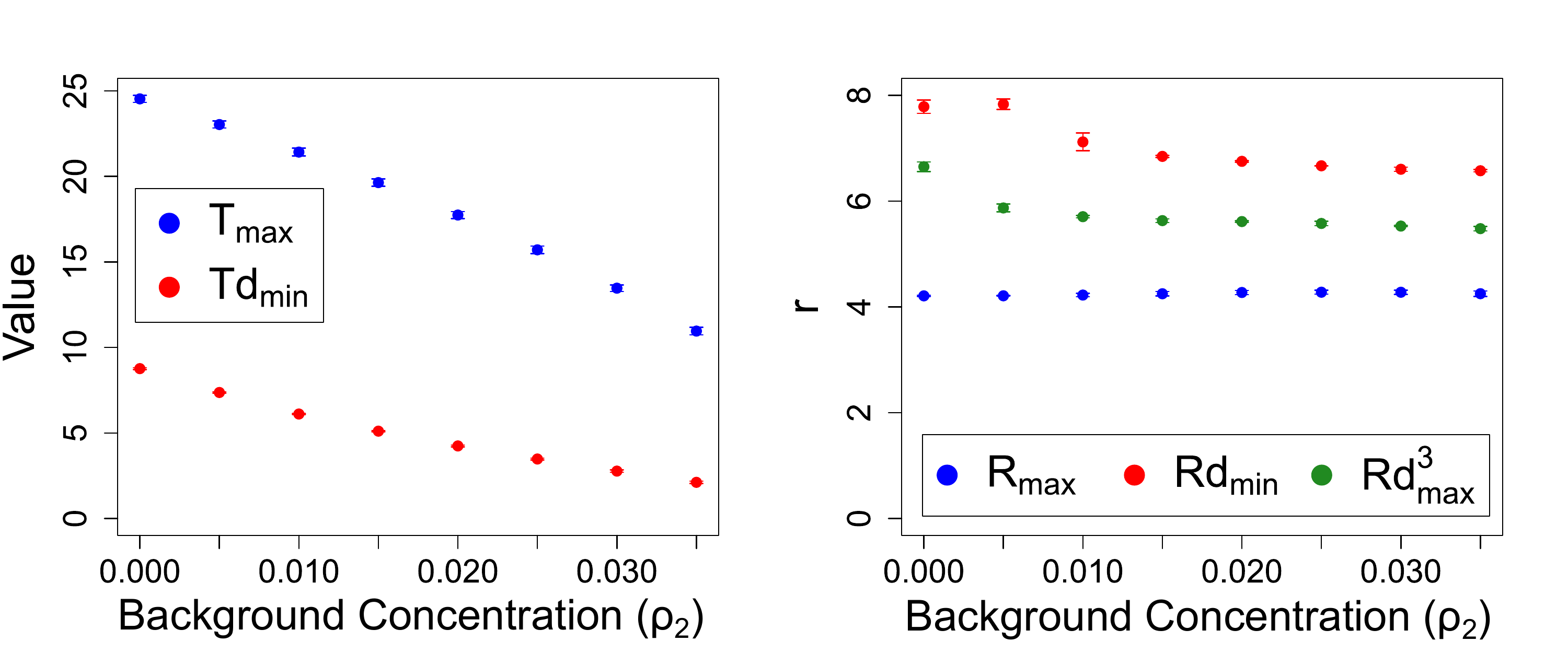}
\caption{$T(r)$ metrics measured on simulated clustered data sets while sweeping through a range of background concentration values. Other simulated cluster parameters are: $\mu_R = 3$ arb., $\rho_1 = 1$, $\beta = 0$, $\xi = 0$.}
\label{series_rho2}
\end{figure}

\begin{figure}[H]
\centering
\includegraphics[width = 1\linewidth]{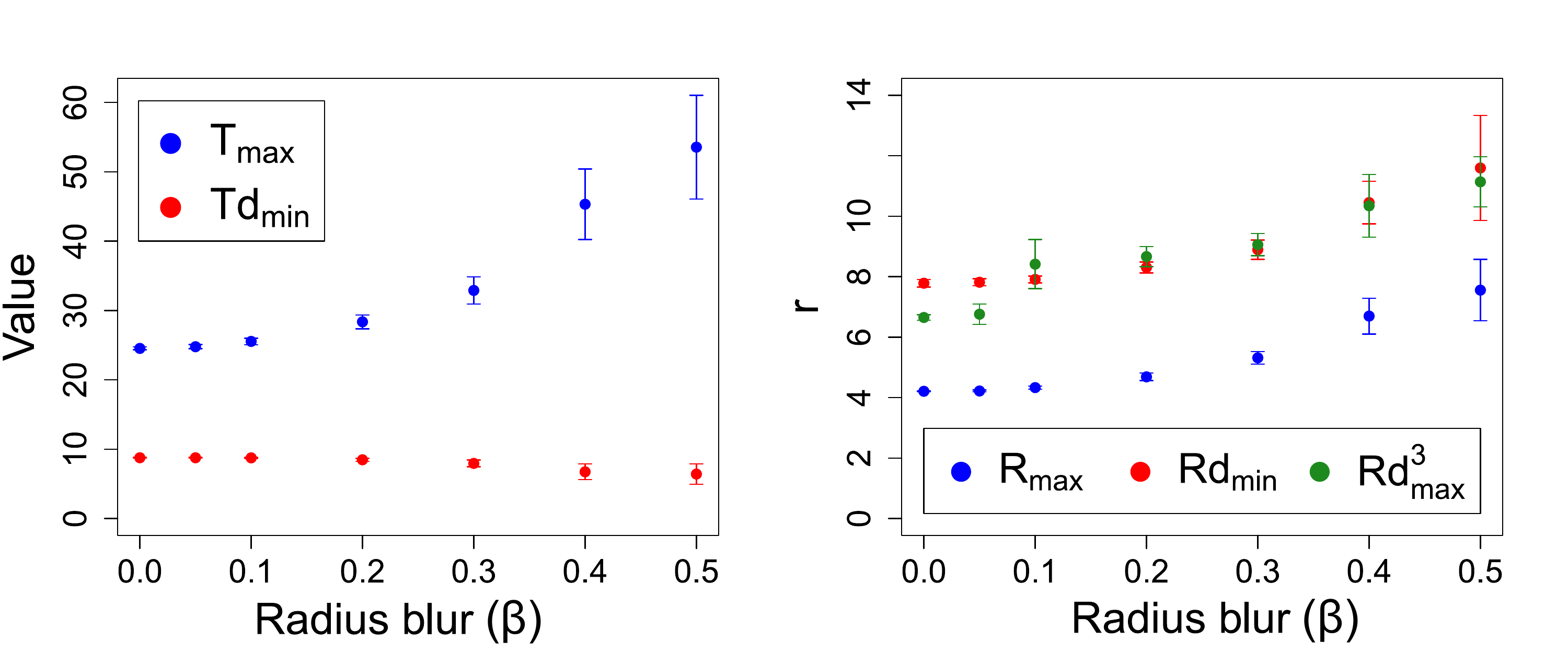}
\caption{$T(r)$ metrics measured on simulated clustered data sets while sweeping through a range of $\beta$ values. Other simulated cluster parameters are: $\mu_R = 3$ arb., $\rho_1 = 1$, $\rho_2 = 0$, $\xi = 0$.}
\label{series_rb}
\end{figure}

\begin{figure}[H]
\centering
\includegraphics[width = 1\linewidth]{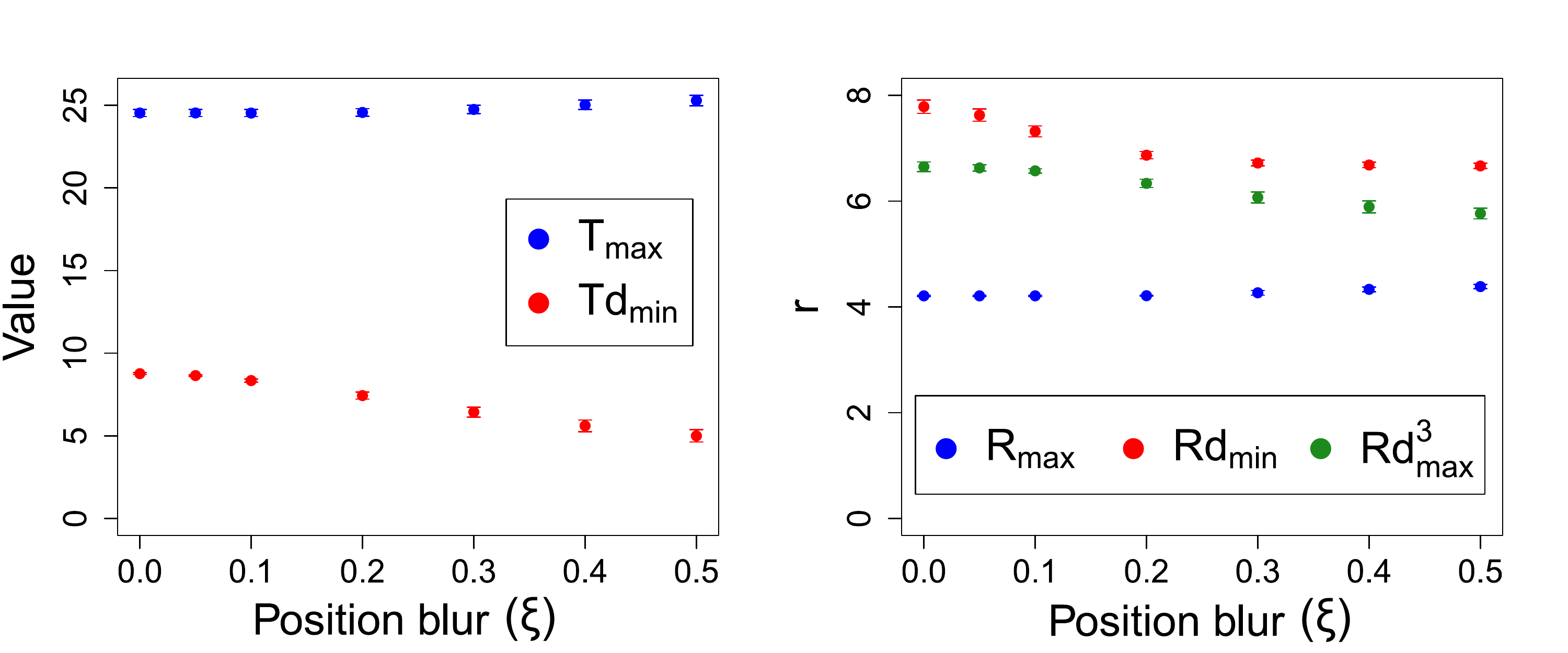}
\caption{$T(r)$ metrics measured on simulated clustered data sets while sweeping through a range of $\xi$ values. Other simulated cluster parameters are: $\mu_R = 3$ arb., $\rho_1 = 1$, $\rho_2 = 0$, $\beta = 0$.}
\label{series_pb}
\end{figure}

Figures \ref{series_radius}-\ref{series_pb} show $T(r)$ metrics for individual cluster parameter sweeps. Table \ref{sweep_vals} shows each parameter's values swept through, as well as the constant held values when not being swept through. The figures show nonlinear, correlated relationships.

\begin{table}[H]
\centering
\caption{Individual Cluster Parameter Sweep Values}
\label{sweep_vals}
\begin{tabular}{lcl}
\hline
\multicolumn{1}{c}{Parameter}                                           & \begin{tabular}[c]{@{}c@{}}Constant\\ Value\end{tabular} & \multicolumn{1}{c}{Sweep Values}      \\ \hline
Cluster Radius ($\mu_R$)                                                & 3 arb.                             & \{2, 3, 4, 5, 8\} arb.             \\
Intra-cluster Conc. ($\rho_1$)                                    & 1                                  & \{0.2, 0.3, 0.4, 0.5, 0.6, 0.7, 1\}          \\
Background Conc. ($\rho_2$)                                       & 0                                  & \{0, 0.005, 0.01, 0.015, 0.02, 0.025, 0.03, 0.035\}   \\
Radius Blur ($\beta$)                                  & 0                                  & \{0, 0.05, 0.1, 0.2, 0.3, 0.5, 0.6\}          \\
Position Blur ($\xi$)   & 0                                  & \{0, 0.05, 0.1, 0.2, 0.3, 0.4, 0.5\}      \\ \hline
\end{tabular}
\end{table}

\section{Principal Component Analysis} \label{SI_sec_pca}

Table \ref{PCA_summary} shows the variance summary for principal component analysis performed on the $T(r)$ metrics of the training data set. These PCs were used as uncorrelated predictors in the machine learning models for $R_w$, $\rho_1$, and $\rho_2$. 

\begin{table}[H]
\centering
\caption{PCA Summary}
\label{PCA_summary}
\begin{tabular}{cccccc}
\hline
                       & PC1    & PC2    & PC3    & PC4    & PC5    \\ \hline
Proportion of Variance & 0.8058 & 0.1689 & 0.0200 & 0.0037 & 0.0016 \\
Cumulative Proportion  & 0.8058 & 0.9747 & 0.9948 & 0.9984 & 1.0000 \\ \hline
\end{tabular}
\end{table}

\section{Uniform Radii Simulations} \label{SI_sec_norb}
In this section, we discuss simulations and results similar to those presented in Section \ref{sec_param_estimation} of the main text, but with the simulation parameter $\beta = 0$ for all simulations. For consistency, we discuss models for $R_w$, but by equation (\ref{rw_normal}) of the main text, $R_w = \mu_R$ when $\sigma_R = 0$, so we are really modeling $\mu_R$. 

To train the models for this case, we followed the same procedure outlined in Section \ref{sec_param_estimation} of the main text, with the only change being that $\beta = 0$ for each of the 10,000 parameter sets of the training data and for all 25,000 testing data simulations. 

A summary of the variance explained by the principal components obtained from this new set of training data is shown in Table \ref{PCA_summary_norb}.

\begin{table}[H]
\centering
\caption{PCA Summary for $\beta = 0$ Models}
\label{PCA_summary_norb}
\begin{tabular}{cccccc}
\hline
                       & PC1    & PC2    & PC3    & PC4    & PC5    \\ \hline
Proportion of Variance & 0.8193 & 0.1533 & 0.0245 & 0.0019 & 0.0010 \\
Cumulative Proportion  & 0.8193 & 0.9726 & 0.9971 & 0.9990 & 1.0000 \\ \hline
\end{tabular}
\end{table}

\begin{figure}[H]
    \centering
    \includegraphics[width=1\linewidth]{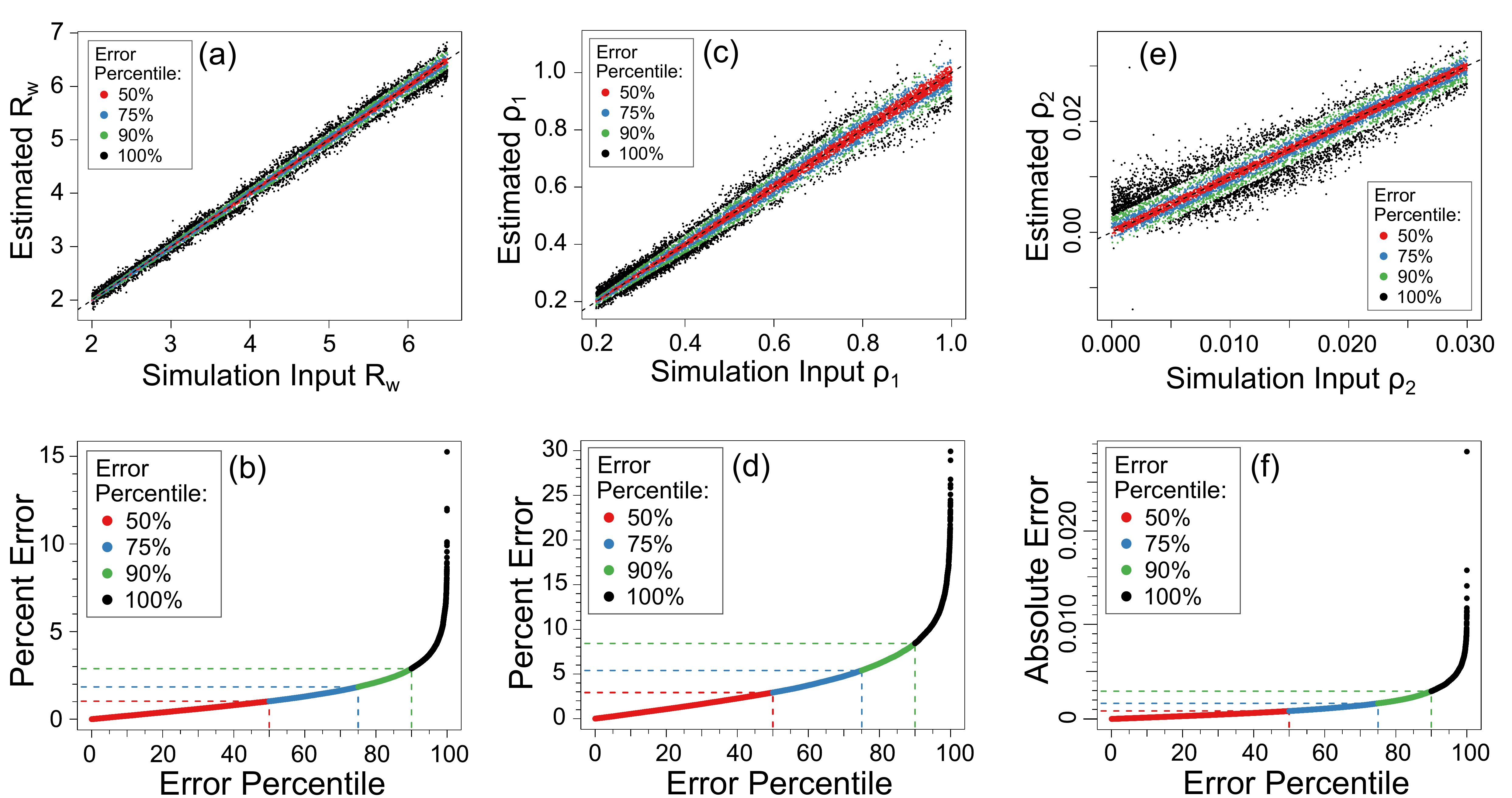}
    \caption{Results from testing data for the BRNN models of weighted radius ($R_w$), intra-cluster concentration ($\rho_1$), and background concentration ($\rho_2$); (a, c, e) True simulated values versus corresponding model estimates for $R_w$, $\rho_1$, and $\rho_2$, respectively. Colors correspond to different error percentiles of the estimates (percent error for $R_w$ and $\rho_1$, absolute error for $\rho_2$); (b, d, f) Ordered error in testing data estimates for $R_w$, $\rho_1$, and $\rho_2$, respectively.}
    \label{mlmodels_norb}
\end{figure}

The RMSEP values of each model from testing with the testing data set are shown in Table \ref{model_RMSEP_norb}, showing that the BRNN model is best for all three parameters. Figure \ref{mlmodels_norb} summarizes the performance of these BRNN models on the testing data set; \ref{mlmodels_norb}(a) shows the true simulated values of $R_w$ versus the corresponding model estimates. The colors correspond to the percent error percentiles of these estimates (i.e. the 50\% of estimates with lowest percent error are shown in red, etc.). Figure \ref{mlmodels_norb}(b) plots the percent error of each estimate sorted in ascending order. This plot shows what percent of model estimates fall below a certain percent error (e.g. we can see that approximately 50\% of estimates have lower that 1\% error). Figures \ref{mlmodels_norb}(c,d) show similar plots for the $\rho_1$ test data estimates from the BRNN model. Figures \ref{mlmodels_norb}(e,f) show similar plots for the $\rho_2$ test data estimates from the BRNN model, except that absolute error is used in place of percent error due to the fact that $\rho_2 \approx 0$ in many of the simulated point patterns.

For these simulated point patterns with $\beta = 0$, the models have increased predictive power from the simulations with non-zero $\beta$. For the $R_w$ model here, 90\% of estimates fall below 3\% error; for the $\rho_1$ model, 90\% of estimates fall below 9\% error; and for the $\rho_2$ model, 90\% of estimates fall below 0.003 (absolute) of the true parameter value. 

\begin{table}[H]
\centering
\caption{RMSEP for Different Models and Parameters}
\label{model_RMSEP_norb}
\begin{tabular}{cllll}
\hline
\multicolumn{1}{l}{}   & Model & $R_w$  & $\rho_1$ & $\rho_2$ \\ \hline
\multirow{3}{*}{RMSEP} & GLM   & 0.1359 & 0.0630   & 0.00347  \\
                       & BRNN  & 0.0781 & 0.0269   & 0.00183  \\
                       & RF    & 0.0862 & 0.0280   & 0.00184  \\ \hline
\end{tabular}
\end{table}

\section{Addressing Assumptions About the Experimental APT Reconstruction} \label{SI_sec_assumptions}

\begin{figure}[H]
    \centering
    \includegraphics[width=0.5\linewidth]{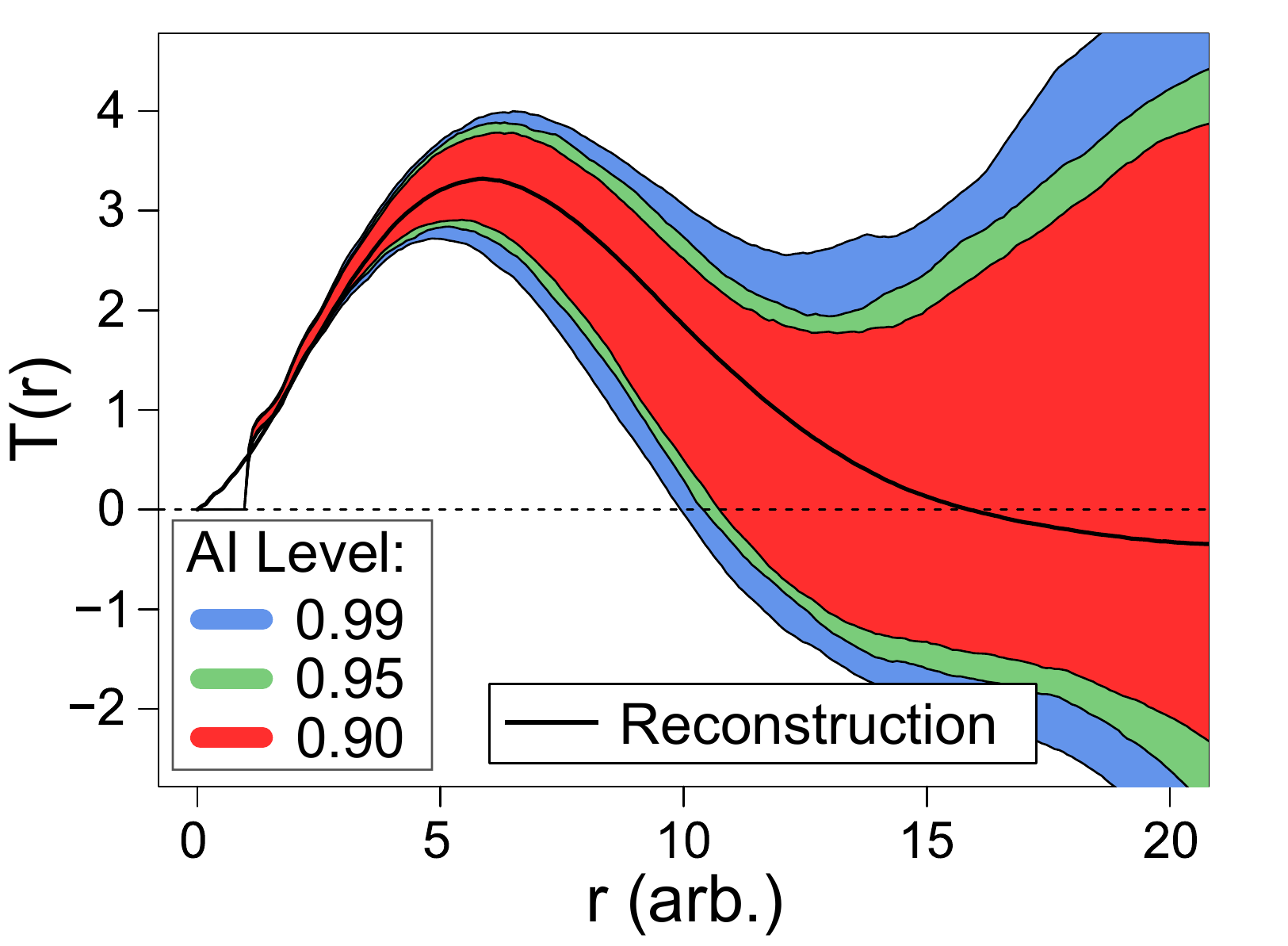}
    \caption{Measured $T(r)$ from scaled reconstruction of AlMgZn alloy compared with level 0.99, 0.95, and 0.90 AI envelopes for simulated clustered data with cluster species fraction $\eta = 0.0511$, $\mu_R = 4.5$, $\beta = 0.238$ ($R_w = 5.192$), $\rho_1 = 0.212$, $\rho_2 = 0.03$, and $\xi = 0.20$.}
    \label{t_sim_vs_real}
\end{figure}

 We now address the assumptions we made to arrive at the estimates shown in Table \ref{apt_estimates} of the main text; that MgZn clusters are spherical, have normally distributed radii, and have uniform intra-cluster concentration. Direct validation of these assumptions would require detailed information about the properties of the clusters in the reconstruction, which are unknown. Instead, we can indirectly corroborate the assumptions by comparing $T(r)$ measured on the scaled reconstruction with $T(r)$ measured on simulated point patterns which do satisfy all of the assumptions. We calculated $T(r)$ for an ensemble of 500 simulated point patterns, each with cluster species fraction $\eta = 0.0511$ and parameters $\mu_R = 4.5$, $\beta = 0.238$ (this defines $R_w = 5.192$), $\rho_1 = 0.212$, $\rho_2 = 0.03$, and $\xi = 0.20$. These parameters were selected to match the parameters estimated for the scaled reconstruction shown in Table \ref{apt_estimates} of the main text, except for $\rho_2$ and $\xi$, which were selected in an ad hoc manner to best match the experimental data, as estimates for their values do not exist. Figure \ref{t_sim_vs_real} shows a comparison of the $T(r)$ curve from the scaled reconstruction with the level 0.99, 0.95, and 0.90 AI envelopes created from this ensemble of 500 $T(r)$ curves. The reconstruction $T(r)$ curve falls within each of these envelopes for all but small radii, which can be explained by the fact that the RCP UPP used in the simulations make it so points cannot fall within some minimum distance of one another, while the APT reconstruction is not constrained to this requirement. Because our assumptions are more concerned with the larger-scale clustering structure of the reconstruction, we can ignore this small simulation artifact in $T(r)$ and focus on larger radius values. Because the reconstruction $T(r)$ falls within the AI envelopes of the simulations, there is no evidence that the structure of the reconstruction is significantly different than the structure of the simulations. We thus conclude that there is no evidence to the contrary of the assumptions made to obtain the estimates in Table \ref{apt_estimates} of the main text; that the clusters within the experimental reconstruction are consistent with those in the simulated training data. 
 
\section{Comparison to MSA} \label{SI_sec_msa}
\begin{figure}[H]
    \centering
    \includegraphics[width=0.5\linewidth]{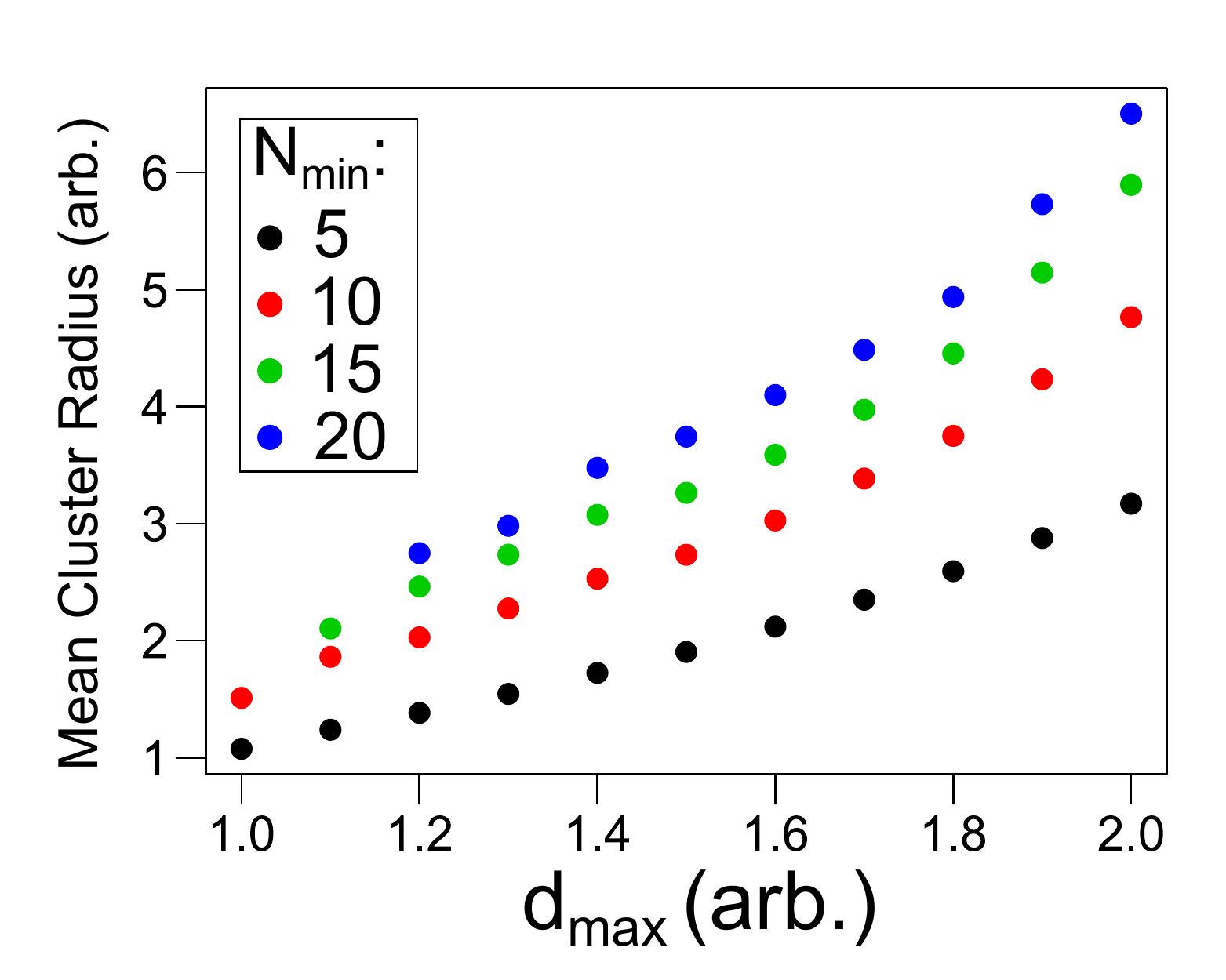}
    \caption{Mean cluster radius of MgZn clusters within the scaled AlMZng alloy APT reconstruction, calculated using the maximum separation algorithm, shown as a function of $d_\text{max}$ and $N_\text{min}$.}
    \label{msa_r}
\end{figure}

Figure \ref{msa_r} shows the mean MSA estimates of mean cluster radius $R_c$ from the scaled APT reconstruction of the AlMgZn alloy as a function of user-input parameters $d_\text{max}$ and $N_\text{min}$. 

\newpage

\bibliography{bibliography_SI}
\bibliographystyle{unsrt}

\makeatletter\@input{xxSI.tex}\makeatother